\documentclass[%
aps,
prx,
reprint,
superscriptaddress,
nofootinbib,
amsmath,
amssymb
]{revtex4-2}
\usepackage[T1]{fontenc}
\UseRawInputEncoding
\usepackage{
    physics,
    graphicx,
    multirow,
    tabularx,
    mathtools, 
    placeins, 
    nicefrac, 
    hyperref, 
    threeparttable, 
    siunitx,
    enumitem,
    comment,
    stmaryrd,
    amsthm,
    bbm,
    makecell
}

\usepackage[capitalize]{cleveref}
\Crefname{section}{Sec.}{Secs.}

\usepackage[dvipsnames]{xcolor}
\hypersetup{
    colorlinks,
    linkcolor={blue!50!black},
    citecolor={blue!50!black},
    urlcolor={blue!80!black}
}

\usepackage[caption=false]{subfig} 
\graphicspath{{Figures/}}

\NewDocumentCommand{\evalat}{sO{\big}mm}{%
  \IfBooleanTF{#1}
   {\mleft. #3 \mright|_{#4}}
   {#3#2|_{#4}}%
}

\newtheorem*{proof*}{Proof}

\begin{document}


\title{Towards early fault tolerance on a 2$\times$N array of qubits equipped with shuttling}

\newcommand{\qmaddress}{\affiliation{Quantum Motion, 9 Sterling Way, London N7 9HJ, United Kingdom}}
\newcommand{\oxddress}{\affiliation{Department of Materials, University of Oxford, Parks Road, Oxford OX1 3PH, United Kingdom}}
\newcommand{\ucladdress}{\affiliation{Department of Physics and Astronomy, University College London, Gower St, London WC1E 6BT, United Kingdom}}
\author{Adam Siegel}
\email{adam.siegel@materials.ox.ac.uk}
\oxddress
\qmaddress

\author{Armands Strikis}
\qmaddress
\ucladdress

\author{Michael Fogarty}
\qmaddress

\date{\today}

\begin{abstract}
It is well understood that a two-dimensional grid of locally-interacting qubits is a promising platform for achieving fault tolerant quantum computing. However in the near-future, it may prove less challenging to develop lower dimensional structures. In this paper, we show that such constrained architectures can also support fault tolerance; specifically we explore a 2$\times$N array of qubits where the interactions between non-neighbouring qubits are enabled by shuttling the logical information along the rows of the array. Despite the apparent constraints of this setup, we demonstrate that error correction is possible and identify the classes of codes that are naturally suited to this platform. Focusing on silicon spin qubits as a practical example of qubits believed to meet our requirements, we provide a protocol for achieving full universal quantum computation with the surface code, while also addressing the additional constraints that are specific to a silicon spin qubit device. Through numerical simulations, we evaluate the performance of this architecture using a realistic noise model, demonstrating that both surface code and more complex qLDPC codes efficiently suppress gate and shuttling noise to a level that allows for the execution of quantum algorithms within the classically intractable regime. This work thus brings us one step closer to the execution of quantum algorithms that outperform classical machines.
\end{abstract}

\maketitle

\section{Introduction}

Quantum computing holds the promise of solving tasks that lie beyond the capabilities of classical computers. Nonetheless, their full potential can only be realised by executing deep quantum algorithms that require extremely low logical error rates, below $10^{-10}$ \cite{Fowler_2012, resource_estimation_shor_algo}, thus far from what can directly be achieved on physical devices \cite{google_surface_code}. Quantum error correction promises to bridge this gap by increasing the qubit overhead to build quantum error correcting codes. However necessary, these architectures represent formidable experimental challenges as they require the entanglement of a very large number of qubits and the repeated measurement of a considerable number of operators called stabilisers \cite{Fowler_2012}. To facilitate their physical implementation, practical constraints are often integrated in the code design, such as locality of the interactions and planar structure of the code. The main example of error correcting code respecting these constraints is the surface code \cite{Bravyi_Kitaev_1998, Kitaev_1997}, and its high threshold --- that is the maximum error rate the code can tolerate to exponentially reduce errors --- makes it one of the best candidates for achieving fault tolerance. However, more general quantum low-density parity-check codes have been constructed that demonstrate better performance in finite-size simulations~\cite{Breuckmann_2017,bravyi2023highthreshold,Panteleev_2021,Higgott_2023,Xu_2023, viszlai2023matching}.
Although these codes seem more challenging to realise experimentally due to their long-range interactions, they have the potential to considerably lower the qubit overhead due to their higher rate, \textit{i.e.} the number of encoded logical qubits per physical qubit, and better distance scaling. Consequently, further research has been undertaken to demonstrate the experimental viability of such complex codes, and suggested solutions for the implementation of their long-range connections, either by swapping \cite{pattison2023hierarchical} or displacing the qubits \cite{viszlai2023matching, Bluvstein_2023, Cai_2023}.

It may seem natural to assume the use of fully 2D architectures to embed such codes, possibly with augmentations to enable some longer-range interactions (as in, e.g., \cite{bravyi2023highthreshold}). However, such complex devices  might not be available for some time; their first fault tolerant iteration may be more limited. For example, the size of an array realised on a quantum chip might be constrained in one direction. This motivated research towards the feasibility of 1D or quasi-1D error correction \cite{Li_2018, shaw2022quantum}. Implementing 2D codes in this kind of devices can theoretically be solved by extending the length of the interactions, thereby increasing the non-locality of the operations. Albeit challenging, this can however be tackled in the same way as long-range interactions in the aforementioned qLDPC codes, \textit{i.e.} by swapping or displacing the qubits.

In this paper, we study the feasibility of quantum error correction in the most constrained experimental setup beyond 1D: a 2$\times$N array of qubits. Long-range interactions are implemented by assuming that the qubits can be collectively shuttled along one of the two lines of the array. Several qubit platforms have been shown to be suitable to implement such an operation with high fidelity,
including atom arrays \cite{Bluvstein_2023} or ion traps \cite{Pino_2021}, but spins qubits \cite{Langrock_2023} may be the most promising one for also implementing the proposed 2$\times$N array.
First, we establish a precise framework to determine the classes of codes that can be efficiently implemented in our device. This study is then applied to two specific examples: the surface code and higher-rate qLDPC codes. Regarding the surface code, we demonstrate that universal quantum computation is possible in a practical setup based on silicon spin qubits, despite the additional experimental constraints that the geometry entails. With the help of numerical simulations accounting for additional shuttling errors, we estimate a resource cost that is moderate enough to run meaningful quantum algorithms in the classically intractable regime. As for higher-rate qLDPC codes, we evaluate their performance under the noise processes that are anticipated for relevant devices, and demonstrate that despite their complexity they do give an advantage over the surface code in a noise domain that is practically achievable.

For simplicity our analysis assumes a strictly 2$\times$N array. In practice, given the long device length that would be required for meaningful post-classical tasks, it is reasonable to assume that a real device could incorporate a plurality of junctions -- thus the overall geometry would be a lattice formed of long 2$\times$N sections. The processes we analyse in this paper would then occur along each long section, while the overall algorithm would benefit from the additional routing advantages of the lattice. We return to this possibility in the discussion.

\section{Framework for finding codes within device restrictions} \label{sec:framework}

In this section, we introduce the formalism for finding QEC codes that naturally suit the 2$\times$N architecture. Before describing the architecture in detail, let us note the perspective we take to judge a code's compatibility.

To start with, we require the architectural constraints to permit implementing a quantum circuit that performs repeated QEC cycles of a suitable code. Specifically, the qubit connectivity graph of the architecture should allow one to repeatedly extract syndrome for error correction. In such a graph, the qubits represent vertices and we take it that entangling two-qubit operations can be implemented between qubits that are connected by an edge. On its own this is not a strong requirement. Any QEC code can be implemented on any connectivity graph as long as one of its connected components hosts the required number of qubits. This could be always achieved in principle, for example by applying a sufficient number of SWAP gates.

Therefore, as an addition to the first criterion, we require syndrome extraction to be efficient at some desired scale. That is, we would like to finish a full QEC cycle in some relatively small number of time steps. While currently this criterion is not well defined in the mathematical sense, it is easier to further specify it after introducing the architecture.

\subsection{Architecture}

\begin{figure}
    \centering
    \includegraphics[width=.8\linewidth]{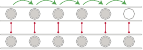}
    \caption{Representation of the 2$\times$N architecture. The device is characterised by two parallel rails of evenly-spaced qubits (gray circles with gray disks inside). Adjacent qubits from different rails are allowed to interact via two-qubit gate operations (red vertical lines with disks on their ends). Finally, the qubits of the first row are allowed to shuttle along their rail (green arrows). Some empty locations are kept at the end of the device to leave space for the qubits to shuttle (empty gray circle).}
    \label{fig:device_layout}
\end{figure}

Let us describe the architecture while keeping the above device criteria in mind. The 2$\times$N architecture is a quasi-1D array where all qubits (data and ancilla) are placed on one of two parallel rails. Separately these rails are treated as 1D strings of evenly spaced qubits, however, an important feature is that the adjacent qubits from separate rails are allowed to interact ``across the ridge'' to perform two-qubit operations such as controlled-Z (CZ) and controlled-X (CNOT), see Fig.~\ref{fig:device_layout}. Clearly such a structure might support other interactions, for example qubits in one or both rails may be able to interact with nearest neighbour qubits within the same rail. However, controlling such interactions might imply additional engineering cost, and in fact such capabilities are {\em not} required in the approaches we describe.

So far the qubit connectivity graph for our 2$\times$N architecture is very limited. To make this more suitable for a wide range of quantum circuits and codes, we introduce the second feature of the architecture - qubit shuttling. Shuttling will play a crucial role in generating different kinds of connectivity graphs that can be used for QEC. Since we are dealing with 1D rails of qubits, we can constrain the shuttling to a `conveyor-belt' model where the whole rail of qubits simultaneously moves along the rail in one direction or the other. After the move, the qubits of one rail may again interact with adjacent qubits in the other rail. Since only the respective qubit positioning of one rail to the other is important, we can always fix one of the rails to stay static.

In this way, the rail of qubits can be shuttled multiple times during a single QEC cycle to generate the desired connectivity graph for syndrome extraction. Here, we consider the canonical syndrome extraction method where ancilla qubits are coupled to the data qubits of the code. If there are no other interactions, the connectivity graph is bipartite --- its exact structure depends on the specific choice of the code. However, by locating all ancilla qubits on one rail and all data qubits on the other, any bipartite connectivity graph (corresponding to an arbitrary code) can be generated using a sufficient number of (possibly long) shuttles. As noted above, this generally would not be practical, and therefore, for each QEC cycle we either set the numbe<r of shuttles (of any length) to be some small constant number that does not scale with the code size, or we restrict the overall shuttling distance (without constraining the number of shuttles). Due to this, we do not expect to produce asymptotically scalable QEC protocols. The same conclusions are also supported by numerical results presented later in the paper.

Therefore, there seem to be only a limited number of classes of quantum stabiliser codes satisfying these constraints. To describe them, let us formally derive the qubit connectivity graphs that can be obtained with our architecture.

\subsection{Formalism}

Consider the scenario where all ancilla qubits are placed on the static rail and data qubits on the other one - the mobile rail. As we explain later, this layout is preferential for the physical platform we showcase. While we do not consider mixed type rails on which both data and ancilla qubits are placed in some sequence, note that they can be more powerful if additional two-qubit operations are allowed between the neighboring qubits on the same rail. Furthermore, assume that both rails have a number $N$ of qubits and are initially aligned. Then, after indexing each data and ancilla qubit along the rail, we see that the $i$-th ancilla qubit may interact with the $i$-th data qubit. Therefore, the biadjacency matrix of the qubit connectivity graph is a diagonal $N\times N$ matrix $H = \mathrm{diag}(1,1, ... , 1)$. Here, the rows correspond to ancilla qubits and columns to the data qubits. 
If the top rail is shuttled $m$ qubit spaces to either direction, so is the diagonal of $H$ shifted by $m$ spaces to either left or right direction of the principal diagonal. For example, shuttling the data qubit rail one position to the left yields us the following transformation
\begin{equation*}
H_0 = \begin{pmatrix}
1 & 0 & 0 &\dots & 0\\
0 & 1 & 0 &\dots & 0\\
0 & 0 & 1 &\dots & 0\\
\vdots & \vdots & \vdots & \ddots & \vdots\\
0 & 0 & 0 &\dots & 1\\
\end{pmatrix} 
\enspace
\underrightarrow{\rule[-4pt]{0pt}{2pt}{shuttle}}
\enspace
H_1 = \begin{pmatrix}
0 & 1 & 0 &\dots & 0\\
0 & 0 & 1 &\dots & 0\\
\vdots & \vdots & \vdots & \ddots & \vdots\\
0 & 0 & 0 &\dots & 1\\
0 & 0 & 0 &\dots & 0\\
\end{pmatrix} 
\end{equation*}
Finally, if we consider interacting qubits before and after the shuttle, we obtain a biadjacency matrix $H = H_0 + H_1$. This does imply a particular order of applying two-qubit operations, which we ignore until we consider specific QEC protocols. Note that such shuttling on a 2$\times$N architecture does not generate periodicity of diagonals (the bottom-left element of $H_1$ is not a 1). However, it can be created by supplementing the protocol with an another shuttle. The farther away two diagonals are, the longer the shuttle operation needed to go between them. Furthermore, not all elements have to be $1$'s along these diagonals, as some interactions may not be implementable due to limited control, or may not be needed.
For example, the biadjacency matrices
\begin{equation*}
H = \begin{pmatrix}
0 & 0 & 0 & 0 & 1\\
1 & 0 & 0 & 0 & 0\\
0 & 1 & 0 & 0 & 0\\
0 & 0 & 1 & 0 & 0\\
0 & 0 & 0 & 1 & 0
\end{pmatrix} 
\quad
H' = \begin{pmatrix}
0 & 1 & 1 & 0 & 1\\
0 & 0 & 1 & 1 & 0\\
1 & 1 & 0 & 1 & 0\\
0 & 1 & 1 & 0 & 0\\
0 & 0 & 0 & 1 & 0
\end{pmatrix} 
\end{equation*}
have two and five non-trivial (non-zero) diagonals respectively.

So far we have described how to generate a biadjacency matrix describing a bipartite qubit connectivity graph. Let us now link this back to finding compatible QEC codes as follows. 
Consider treating ancilla qubits as parity checks that interact with their respective data qubits during the shuttling phase and afterwards are measured out in some basis. Then the generated biadjacency matrix $H$ describes a potential parity check matrix of some code $\mathcal{C}$ whose QEC cycle can be implemented in a number of steps proportional to the number of shuttles. The same procedure of shuttling, interacting and measuring can be repeated over again, and therefore multiple QEC cycles can be continuously performed. 
There are still some degrees of freedom left. For example, we have not specified what kind of Pauli string each parity check (row of $H$) is. Therefore, any stabiliser code whose parity check matrix can be written with non-zero elements matching the $1$'s of the generated $H$ would count as a suitable code --- as long as the number of shuttles or their overall length is small. In fact, that is the approach we take.
We find codes with parity check matrices that have few non-trivial diagonals (low number of shuttles) or all of them are close to the main diagonal (short shuttling distance). We then map them to our 2$\times$N architecture.
One can also go the other way around and try to design a stabiliser code by first generating an arbitrary $H$ and assigning rows to specific parity checks. However, knowing which Pauli strings to choose to create a good code with commuting stabilisers seems to be a difficult task.

Let us again note that sequencing the entangling operations between data and ancilla qubits can be crucial when it comes to extracting syndrome. By only considering CSS codes, we can always perform all $Z$ type stabilisers while we shuttle the data qubits in one direction and all $X$ type stabilisers while we shuttle them back to their initial position. While this increases the total number of shuttles and the qubit idling time, the overall shuttling distance is kept the same as when both stabiliser measurements are interleaved. Such sequencing is used later when we consider more general qLDPC codes. Unfortunately, this method does not guarantee that hook errors reducing the code's effective distance will not arise; these need to be considered separately.

Our proposed recipe is thus to find parity check matrices with a few or closely located non-trivial diagonals, which we demonstrated are suitable for the $2 \times N$ architecture. Let us now provide a couple of examples.

\begin{figure}
    \centering
    \subfloat{
        \centering
        \includegraphics[width=0.5\linewidth]{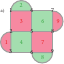}}
    \hfill
    \subfloat{
        \centering
        \includegraphics[width=0.45\linewidth]{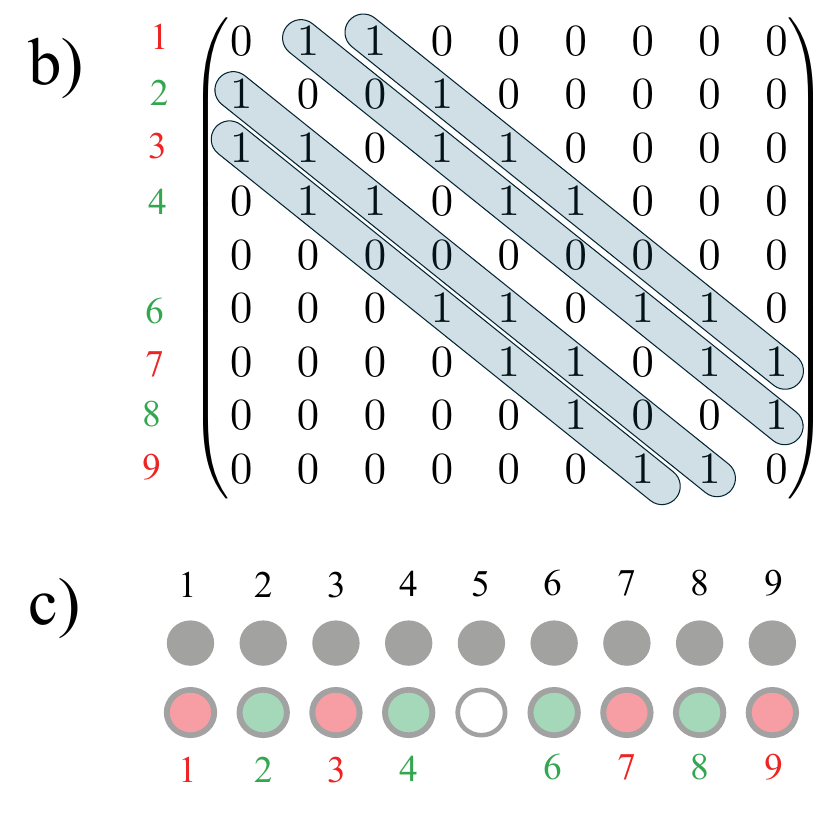}}
    \caption{Three-way mapping for the rotated surface code. (a) The canonical 2D layout, (b) its parity check matrix with an additional trivial stabiliser and (c) a schematic for the layout on a 2$\times$N architecture. Here, red/green colours indicate X/Z stabilisers respectively.} 
    \label{fig:RotSurface}
\end{figure}

\subsection{Code examples} \label{sec:code_examples}

\subsubsection{Rotated surface code}

The first and potentially the most promising example is the rotated surface code. It is usually depicted in a graphical way as in Fig.~\ref{fig:RotSurface}(a). Since for this code there are weight-four stabiliser generators, its parity check matrix has at least four non-trivial diagonals.
In fact, by assigning some parity check qubits to have no support, we can construct a square parity check matrix of a rotated surface code that has exactly $4$ non-trivial diagonals. See Fig.~\ref{fig:RotSurface} for a three-way mapping of a distance-$3$ rotated surface code. As explained before, having four diagonals is nonetheless not sufficient to ensure that a stabiliser cycle can be implemented in four steps - the order of the entangling gates is highly significant. We however verify in Appendix \ref{appendix:gate_ordering} that measuring the diagonals of Fig. \ref{fig:RotSurface}(b) in the order 1-4-2-3 (where the diagonals are indexed from left to right) guarantees a minimum number of shuttles, a minimum shuttling distance and no distance-reducing hook errors.

\subsubsection{Hypergraph product codes}

Another interesting class of codes are the hypergraph product (HGP) codes that are composed with a repetition code as one of the seed codes. The other classical code should still be an LDPC code, but can be chosen arbitrarily. For such codes, the number of shuttles required to obtain the full connectivity graph grows with the code size, however, the full shuttling distance scales only as a square-root of the number of qubits. To see this, take $A$ to be an $(n-1) \times n$ parity check matrix of a classical repetition code
\begin{equation*}
A = 
\begin{pmatrix*}
1 & 1 & 0 & 0 & \dots\\
0 & 1 & 1 & 0 &\dots\\
0 & 0 & 1 & 1 &\dots\\
\vdots & \vdots & \vdots & \vdots & \ddots
\end{pmatrix*}
\end{equation*}
and $B$ to be a random sparse $l \times k$ matrix representing some classical LDPC code. Then the HGP code of $A$ and $B$ has parity check matrices
\begin{align*}
H_x & = 
\begin{pmatrix*}
    A \otimes  \mathbbm{I}_k, \mathbbm{I}_{n-1} \otimes B^T
\end{pmatrix*}\\
H_z & =
\begin{pmatrix*}
    \mathbbm{I}_n \otimes B, A^T \otimes \mathbbm{I}_l 
\end{pmatrix*},
\end{align*}
where $\mathbbm{I}_m$ is an identity matrix of size $m \times m$. 
Since one has the freedom to index the data qubits in any order (i.e. one can switch columns around) it is easy to rearrange $H_x$ in the block form as
\begin{equation*}
H_x = 
\begin{pmatrix*}
\mathbbm{I}_k & B^T &\mathbbm{I}_k & 0 & 0 & 0 & 0  & \dots \\
0 & 0 &\mathbbm{I}_k & B^T &\mathbbm{I}_k & 0 & 0 &  \dots \\
0 & 0 & 0 & 0 &\mathbbm{I}_k & B^T &\mathbbm{I}_k & \dots \\
\vdots & \vdots & \vdots & \vdots & \vdots & \vdots & \vdots & \ddots &
\end{pmatrix*}.
\end{equation*}
Similarly, one can rearrange $H_z$ to have the non-trivial diagonals closer to the principal diagonal,
\begin{equation*}
H_z = 
\begin{pmatrix*}
B & \mathbbm{I}_l & 0 & 0 & 0 & 0  & \dots \\
0 & \mathbbm{I}_l & B &\mathbbm{I}_l& 0 & 0 &  \dots \\
0 & 0 & 0 &\mathbbm{I}_l & B &\mathbbm{I}_l & \dots \\
\vdots & \vdots & \vdots & \vdots & \vdots &\vdots & \ddots &
\end{pmatrix*}.
\end{equation*}
Finally, the total parity check matrix $H$ is rearranged by placing block rows of $H_z$ and $H_x$ in an alternating fashion (as for the rotated surface code presented above) resulting in
\begin{equation*}
H = 
\begin{pmatrix*}
B & \mathbbm{I}_l & 0 & 0 & 0 & 0  & 0  &\dots \\
\mathbbm{I}_k & B^T &\mathbbm{I}_k & 0 & 0 & 0 & 0  & \dots \\
0 & \mathbbm{I}_l & B &\mathbbm{I}_l& 0 & 0 & 0  & \dots \\
0 & 0 &\mathbbm{I}_k & B^T &\mathbbm{I}_k & 0 & 0 &  \dots \\
0 & 0 & 0 &\mathbbm{I}_l & B &\mathbbm{I}_l & 0  &\dots \\
0 & 0 & 0 & 0 &\mathbbm{I}_k & B^T &\mathbbm{I}_k & \dots \\
\vdots & \vdots & \vdots & \vdots & \vdots & \vdots & \vdots &\ddots &
\end{pmatrix*},
\end{equation*}
whose non-trivial diagonals are at most $O(\sqrt{N})$ data qubits apart from each other. This represents the shuttling distance that is necessary to generate the full connectivity graph.
Note that, since we take $B$ to correspond to an arbitrary LDPC code, the generation of most non-trivial diagonals involves only $O(\sqrt{N})$ qubits. The rest of the qubits are idle, and hence one needs to account for the potential negative effects of idling on the code's performance in practice (see Section \ref{sec:hgp}).

\subsubsection{Generalised bicycle codes}

Finally, consider two-block LDPC codes whose parity check matrices take the form of 
\begin{align}
H_x & = 
\label{eq:hx1}
\begin{pmatrix*}
    C, \enspace D
\end{pmatrix*}\\
H_z & =
\label{eq:hz1}
\begin{pmatrix*}
    D^T, C^T
\end{pmatrix*}
\end{align}
where both $C$, $D$ are sparse binary matrices that commute. A way to ensure this commutation is to take $C$ and $D$ as any two $l \times l$ sparse circulant matrices that have the form
\begin{equation*}
C = 
\begin{pmatrix*}
a_0 & a_{l-1} & a_{l-2}  & \dots & a_1\\
a_1 & a_0 & a_{l-1}  &\dots & a_2\\
a_2 & a_1 & a_0  &\dots & a_3\\
a_3 & a_2 & a_1 &\dots & a_4\\
\vdots & \vdots & \vdots & \vdots & \vdots
\end{pmatrix*},
\end{equation*}
where $a_i \in \mathbbm{F}(2)$ such that only a constant number of $a_i$ are non-trivial. These matrices have an intrinsically diagonal structure, therefore, the number of shuttles required to generate the qubit connectivity graph for the corresponding parity check matrices (Eqn.~\ref{eq:hx1},~\ref{eq:hz1}) is constant. Codes constructed with such sparse circulant matrices have been shown to perform well within the code capacity-based simulations~\cite{Panteleev_2021}, therefore, it is interesting to consider them in our 2$\times$N architecture where shuttling noise has to be accounted for. While the number of shuttles is constant, note that, in general, the shuttling distance for these codes scale with the size of the code block.

In the following, we present realistic noise simulations of the mentioned codes under the 2$\times$N architecture. We focus on the rotated surface code for universal fault tolerant quantum computation and later discuss how codes from other classes can be used for efficient quantum memory.

\section{Universal quantum computation on a spin-qubit surface code} \label{sec:universal_quantum_computation}

The previous section discussed how to embed an individual code block on a 2$\times$N device by taking advantage of shuttling. This is only the first step towards the implementation of a fully functional quantum computer, as we still need to discuss how operations between logical qubits can be implemented despite the strong architectural constraints. In this section, we will demonstrate that universal quantum computation is indeed possible when embedding surface-code-like error correcting codes in a $2 \times N$ device equipped with shuttling. In order to ensure that our proposition is practical from the experimental point of view, we will focus on a silicon spin qubit device. Shuttling has been demonstrated to be implementable at high fidelity on such a platform \cite{Yoneda_2021, Seidler_2022}. This motivates the introduction of additional constraints discussed below.

\subsection{Silicon spin qubits}

Electron spins in silicon quantum dots (QDs) are a promising physical platform to perform quantum computing: single- and two-qubit gates with fidelity well above $99\%$ \cite{Xue_2022, Noiri_2022, Mills_2022} or even beyond $99.9\%$ \cite{Yoneda_2017} have been demonstrated, simple instances of quantum error correction have been shown to be feasible \cite{O_Gorman_2016} or have already been implemented \cite{Takeda_2022}, and the technology has been scaled to processors with up to 6 qubits \cite{Philips_2022}. Furthermore, the compatibility with advanced manufacturing techniques \cite{Maurand_2016, Zwerver_2022} and cryogenic classical electronics \cite{XueXiao2021Ccco, ruffino2021integrated} make them ideal candidates for large scale integration \cite{Gonzalez-ZalbaM.F.2021Ssqc}.
 
While silicon architectures are expected to eventually provide dense two-dimensional grids of qubits \cite{Veldhorst_2017, Boter_2022}, early silicon quantum processor designs are particularly interesting since they could meet the requirements for the implementation of the protocol described in this paper. These are $(i)$ a bilinear qubit topology, $(ii)$ information shuttling. Both these criteria have been met experimentally: devices with 2$\times$N arrays of QDs have been demonstrated \cite{hutin2019gate}; and information transfer has been achieved using spin shuttling techniques in the form of bucket brigade \cite{Yoneda_2021} or conveyor belt approaches \cite{Seidler_2022}.
The first method makes use of tunneling to shuttle electrons, by successively lowering the potential of subsequent quantum dots along their way to generate their movement. On the contrary, the second method aims at always keeping the wavefunction at a minimum of a smoothly moving sine wave, without tunneling out of it (see Fig. 2 of \cite{Langrock_2023}). In our paper, as we aim to shuttle all data qubits collectively, the conveyor-belt approach is advantageous: indeed electrons can just be placed at the minima of the moving sine wave potential. Only four signals are thus required for the shuttling of all the data qubits \cite{Langrock_2023}. The drawback is that each electron must occupy the space of four clavier gates rather than one.

There is of course a natural set of gates that one can directly implement with a class of silicon spin qubit device (and here we assume single-spin representations of qubits). More precisely, local phase rotations \cite{Camenzind_2022} and two-qubit operations such as C-Phase gates \cite{Xue_2019} have been shown. However, general single-qubit rotations can be more challenging to localize, and may be more naturally realised on a wide scale, via the interaction of the spins with a global oscillatory field \cite{patomäki2023pipeline}.

\begin{figure}
    \includegraphics[width=\linewidth]{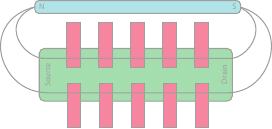}
    \centering
    \caption{Schematic representation of a device where the $g$-factors in the first row are all higher than the $g$-factors in the second row. The green rectangle represents the source and drain, while the red and blue ones respectively represent the gates confining the electrons and the micromagnet. Two field lines are additionally drawn, showing the difference in magnetic field between both rows of the 2$\times$N array. This is in turn responsible for a difference in Zeeman splitting, which can be utilised to selectively target either the spins of the first row or those of the second row.}
    \label{fig:micromagnet}
\end{figure}

For the present analysis, we will require only a modest refinement of unconditionally-global Hadamard gates. It will suffice to apply the operation to the entirety of one of the two linear arrays of the processor. This partial global, or `semi-global' operation could be implemented in silicon devices using frequency engineering, for example by placing a micromagnet next to the array \cite{Kawakami_2014} or by using magnetic materials in the gate stack on one side of the array \cite{CrawfordO.2023Cass}.
Such approaches could create a frequency difference between the two linear arrays that can be selectively addressed with known electron spin resonance techniques \cite{Vahapoglu_2021}. A schematic implementation of such frequency difference between the two lines of the array is pictured in Fig. \ref{fig:micromagnet}, via the use of a micromagnet. Operating at a global magnetic field $B_0=1.4$T, one can expect the $g$-factor dispersion in the device to be around $\Delta f=60$MHz \cite{Jones_2018} In order to sufficiently separate the frequencies of both lines of the array, one could therefore set the micromagnet-induced magnetic field difference to $\Delta B=5h\Delta f/g\mu_B=10$mT. Here $h$ is Planck's constant, $g=2$ is the electron's Land\'e factor and $\mu_B$ is Bohr's magneton. Given a spacing of 100nm between the lines, this would mean a gradient of 0.1mT/nm, which is comfortably below demonstrated gradients, around 0.8mT/nm \cite{KlemtBernhard2023Emoa}.

In our setup, the first and second one-dimensional arrays contain the data and ancilla qubits, respectively and the processor is subject to a magnetic field to polarise the spins. Each data qubit is encoded by a single quantum dot, initialised in the $\ket{0}$ state (spin down) at the beginning of the computation, for example by spin relaxation. On the contrary, each ancilla qubit is encoded by two electrons in a singlet or triplet state. This is consistent with established techniques for preparation and measurement: at the start of each stabiliser cycle, the ancilla qubits are initialised in the singlet state by first, applying a differential gate potential between QDs to produce an energy detuning that will relax the system in the singlet (02) [or (20)] state, followed by a ramp to the singlet (11) state (adiabatic with respect to the singlet anticrossing and diabatic with respect to the singlet-triplet anticrossing). At the end of the cycle, the stabiliser is measured by projecting the pair of spins to one of the QDs which, through Pauli spin blockade reveals a different measurement outcome for singlet or triplet states \cite{Betz_2015, West_2019}.
Finally, note that this choice of singlet-triplets requires the occupation of two quantum dots per ancilla qubit. Besides, measurement apparatuses cannot be included in the data row as they would block the shuttling, hence can only exist in the ancilla row. For these reasons, the ancilla row has to be denser than the data row. This additional space between data qubits is not going to waste as it can be used to fit in the four clavier gates per electron required by the conveyor-belt mode of shuttling.

\subsection{Logical gates and qubit layout}

One of the major constraints discussed above is the absence of local single-qubit gates (apart from phase gates). As such, if one wants to implement, say, a logical $X$ gate on a given logical qubit, applying transversal $X$ gates on the corresponding surface code would not meet our constraints, as these gates would have to be applied on all data qubits in the device, thereby implementing a global logical $X$. Local transversal single-logical-qubit gates are thus not permitted in our device (apart from $Z$ gates). This motivates the choice of a protocol that would not make use of such transversal gates, but rather, of lattice surgery only. Indeed, this operation can be made local as our device does feature selective $CZ$ gates. One such approach is described in \cite{Litinski_2018, Litinski_2019}: using Clifford+$T$ as a universal gate set, these papers show that Clifford gates can be commuted through $T$ gates and incorporated in later measurements, thereby only leaving multi-qubit Pauli measurements and multi-qubit $\pi/8$ rotations to implement (Fig. 4 in \cite{Litinski_2019}).
These rotations can in turn be performed by consuming a magic state through additional multi-qubit measurements. This protocol thus removes the necessity to implement local Clifford gates that are impractical in our device. However, it comes at the cost of complexified measurements which require a modified lattice surgery protocol. As for the magic state, it can be prepared via state injection and subsequent magic state distillation. The latter only requires the implementation of logical CNOTs, which can equally be performed via lattice surgery. Using this scheme, the remaining operations our device must be able to implement for universal quantum computation are thus: $(i)$ stabiliser measurements, $(ii)$ all variants of lattice surgery as described in \cite{Litinski_2018} and $(iii)$ state injection. An implementation of these three components within our architectural limitations is described in Section \ref{sec: circuits}.

Before discussing the circuit-level implementation of these elements, it is first relevant to discuss the layout that we adopt for our logical qubits, as this will have an impact on the ordering of the gates in each circuit (\textit{e.g.} to avoid distance-reducing hook errors).

\begin{figure}
    \centering
    \includegraphics[width=\linewidth]{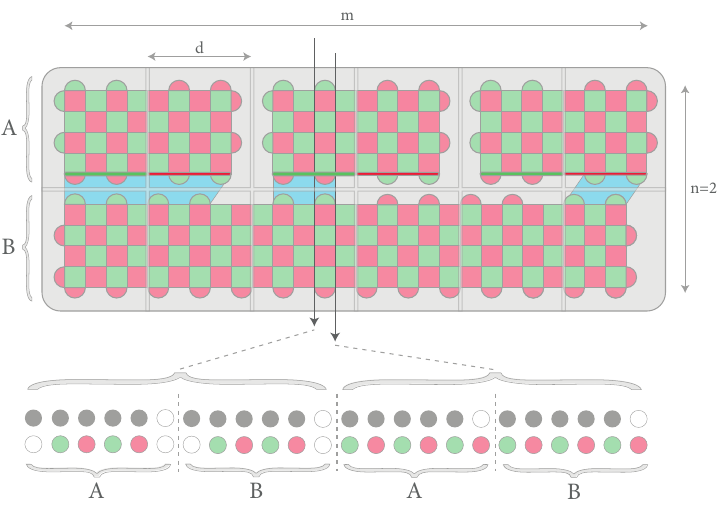}
    \caption{Qubit layout for universal quantum computation on the 2$\times$N architecture. We choose to arrange the logical qubits on a grid of $n\times m$ square tiles of size $d\times d$, where $d$ is the code distance \cite{Litinski_2019}. We set $n=2$ to minimise the shuttling distance. $X$ and $Z$ stabilisers are respectively represented with red and green squares or half-disks. The direction of shuttle is indicated by the vertical dark gray arrows. The top half (region A) contains the logical qubits storing the logical information, while the bottom half (region B) is a logical ancilla bus that can be used to perform long-range interactions between the logical qubits. Each logical qubit is a rectangular patch of surface code, whose $X$ and $Z$ logical operators are represented by red and green lines. Both these logical operators are adjacent to region B, so that they can interact with the logical ancilla. In this example, the ancilla is initialised so as to measure the operator $Y_1\otimes Z_2\otimes X_3$ with three lattice surgeries, represented with the blue boxes (see Fig. 44 of \cite{Litinski_2019} for more details). Below, the portion of the layout corresponding to the qubits along the arrows is linearised, showing the alternation of regions A and B in the 2$\times$N architecture.}
    \label{fig:qubit_layout}
\end{figure}

In order to embed an $n\times m$ grid of $d\times d$ tiles of data qubits (as in \cite{Litinski_2019}) in a 2$\times$N device, one must slice the whole grid, say, in the vertical direction. If we assume for now that each tile represents exactly one surface code, this means two consecutive data qubits within a row of a given patch are now separated by $nd$ data qubits rather than $d$. The consequence is shuttle operations that are $n$ times longer such that these two qubits can participate in the same stabiliser measurement. As shuttling is prone to errors, one way to minimise such long shuttles while permitting full quantum computation is to set $n=2$ and consider a $2\times m$ grid of tiles, where the logical information would only be stored in the first row of the grid (region A), while the second row (region B) would serve as a logical ancilla bus used for long-range interactions between the logical qubits of region A (Fig. \ref{fig:qubit_layout}). Additionally, a magic state factory is included at the right end of the device. In order to enable all types of interactions between the logical qubits of region A, both their logical $X$ and $Z$ operators should be adjacent to region B. This motivates the use of wide qubits as in Fig. \ref{fig:qubit_layout}, on top of an ancilla bus that can be initialised differently depending on the desired operation. The slicing is performed in the vertical direction, meaning that the first row of the 2$\times$N device will contain an alternation of $d$ data qubits from region A and $d$ data qubits from region B. In order to entangle physical data qubits of consecutive columns with a physical ancilla qubit, one must therefore shuttle by roughly $2d$ increments in one direction, and the same distance back (instead of $d$ in the case of a single encoded surface code).

\subsection{Stabilisers implementation} \label{sec: circuits}

\subsubsection{Gate ordering} \label{sec:gate_ordering}

\begin{figure}
    \centering
    \includegraphics[width=\linewidth]{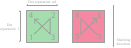}
    \caption{Gate ordering enabling one to measure $X$ and $Z$ stabilisers in an interleaved (but staggered) fashion. While $X$ stabilisers perform steps 5 and 6, $Z$ stabilisers undergo steps 1 and 2 again, and so on. The distance between data qubits (in terms of number of quantum dots) is indicated. This sequencing guarantees a minimum shuttling distance, a minimum number of shuttles and no distance-reducing hook errors.}
    \label{fig:gate_ordering}
\end{figure}

The next step is to ensure that our three building blocks needed for fault tolerant quantum computation --- stabiliser measurements, modified lattice surgery and state injection --- can indeed be implemented within our constraints. Before giving explicit circuits or procedures implementing these operations, one must first determine the order in which data qubits must be entangled within a given stabiliser measurement. This is an important matter as a suboptimal choice of ordering can lead to hook errors which effectively reduce the distance of the code. Additionally, a proper sequencing can help reduce the stabiliser circuit depth, number of shuttles and shuttling distance by interleaving the gates of the $X$ and $Z$ stabilisers. We verify in Appendix \ref{appendix:gate_ordering} that the gate ordering depicted in Fig. \ref{fig:gate_ordering} $(i)$ does not create any hook error; $(ii)$ allows one to interleave the gates of the $X$ and $Z$ stabilisers; $(iii)$ is nearly optimal in terms of the number of shuttles and total shuttling distance.

In the case of modified lattice surgery however, the question of interleaving the dislocations and twists (see next section) with the regular stabilisers is much more complex. Not doing so is always an option, however, that comes at the cost of additional shuttles, and leaves some qubits idle while others are being measured. We leave a study similar to Fig. 16 of \cite{Litinski_2018} to further research.

\subsubsection{Stabiliser circuits with singlet-triplet ancilla qubits}

Given the silicon spin platform which we are principally considering, we take it that the preferred embodiment of an ancilla qubit is via two spins, with measurement occurring by differentiating between singlet and triplet states. After the ancilla is initialised in the singlet state, if there are an odd number of errors affecting the data qubits then the ancilla should transform to a triplet state, while an even number of data qubit errors should leave it in the singlet state. A Pauli spin blockade measurement then allows one to extract the value of the stabiliser.

\begin{figure}
    \subfloat[Z stabiliser]{
        \centering
        \includegraphics[width=0.45\columnwidth]{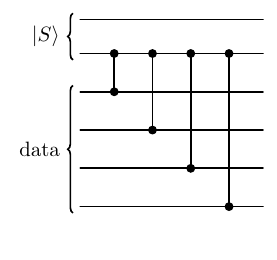}}
    \hfill
    \subfloat[X stabiliser]{
        \centering
        \includegraphics[width=0.45\columnwidth]{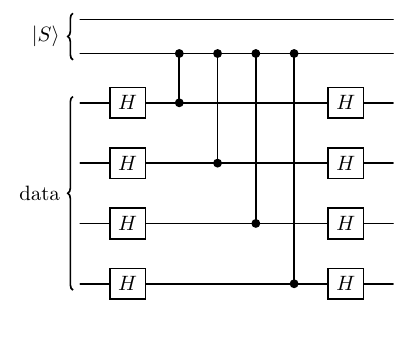}}
    \\
    \subfloat[Dislocation]{
        \centering
        \includegraphics[width=0.45\columnwidth]{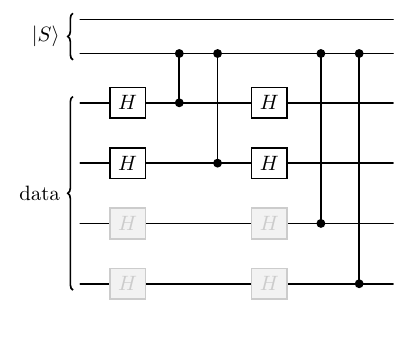}}
    \hfill
    \subfloat[Twist]{
        \centering
        \includegraphics[width=0.45\columnwidth]{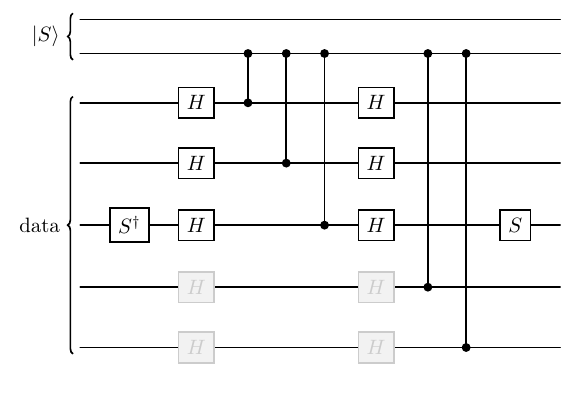}}
    \caption{Circuit implementation of all kinds of stabilisers needed for regular error correction and modified lattice surgery \cite{Litinski_2018}, using a singlet-triplet ancilla qubit. This implementation also respects a semi-global implementation of Hadamard gates, meaning that they are globally applied on all data qubits. The light gray Hadamards are only included for this purpose and cancel each other.} 
    \label{fig:circuit_all_stabs}
\end{figure}

Fig. \ref{fig:circuit_all_stabs} gives a circuit implementation of all types of stabilisers involved in regular error correction as well as lattice surgery, using $CZ$ gates and semi-global Hadamards only (\textit{i.e.} affecting all data qubits at the same time). These stabilisers fall into four categories: regular $X$ stabilisers, regular $Z$ stabilisers, dislocations (stabiliser checks involving both $X$ and $Z$) and twists (stabiliser checks involving $X$, $Z$ and $Y$). The two latter appear in the modified lattice surgery protocols of \cite{Litinski_2019} when $X$ and $Z$ boundaries are facing (rather than both $X$ or both $Z$ in the conventional lattice surgery picture). One can check that, with this implementation, an error on a data qubit always transforms the singlet state $\ket{S}=(\ket{01}-\ket{10})/\sqrt{2}$ into the triplet state $\ket{T}=(\ket{01}+\ket{10})/\sqrt{2}$. This would not be the case if we used CNOTs for the $Z$ stabilisers (with data qubits as controls and one ancilla line as target), as it would transform a singlet state into a different triplet state $\ket{T'}=(\ket{00}-\ket{11})/\sqrt{2}$.

This proves crucial when implementing dislocations and twists. Indeed, as an example, let us consider the measurement of the dislocation operator $XZ$ on data qubits in the $\ket{-}\otimes\ket{1}$ state. This should lead to the final measurement of a singlet state. Let us first assume that the $Z$ parity measurement is implemented with a CNOT targeting one qubit of the ancilla pair (rather than a $CZ$ as in Fig. \ref{fig:circuit_all_stabs}). The quantum state of the data/ancilla system evolves as follows:
\begin{align*}
    \ket{-}\ket{1}(\ket{01}-\ket{10})/\sqrt{2} &\longrightarrow \ket{-}\ket{1}(\ket{01}+\ket{10})/\sqrt{2}\\
        &\longrightarrow \ket{-}\ket{1}(\ket{11}+\ket{00})/\sqrt{2}
\end{align*}
The final ancilla state is a triplet --- this implementation gives the wrong measurement outcome. Let us now assume that the $Z$ parity measurement is implemented with a $CZ$ gate, as prescribed in Fig. \ref{fig:circuit_all_stabs}. This time, the quantum state of the data/ancilla system transforms as follows:
\begin{align*}
    \ket{-}\ket{1}(\ket{01}-\ket{10})/\sqrt{2} &\longrightarrow \ket{-}\ket{1}(\ket{01}+\ket{10})/\sqrt{2}\\
        &\longrightarrow \ket{-}\ket{1}(\ket{01}-\ket{10})/\sqrt{2}
\end{align*}
The ancilla state is thus back to the singlet state, which is the correct behaviour.

The twist measurement can be implemented in a similar way as the dislocation. Also, the presence of the $S$ and $S^\dagger$ gates is not problematic as these are phase gates, which can be implemented locally.

\subsubsection{Full circuit}

\begin{figure}
    \centering
    \includegraphics[width=\linewidth]{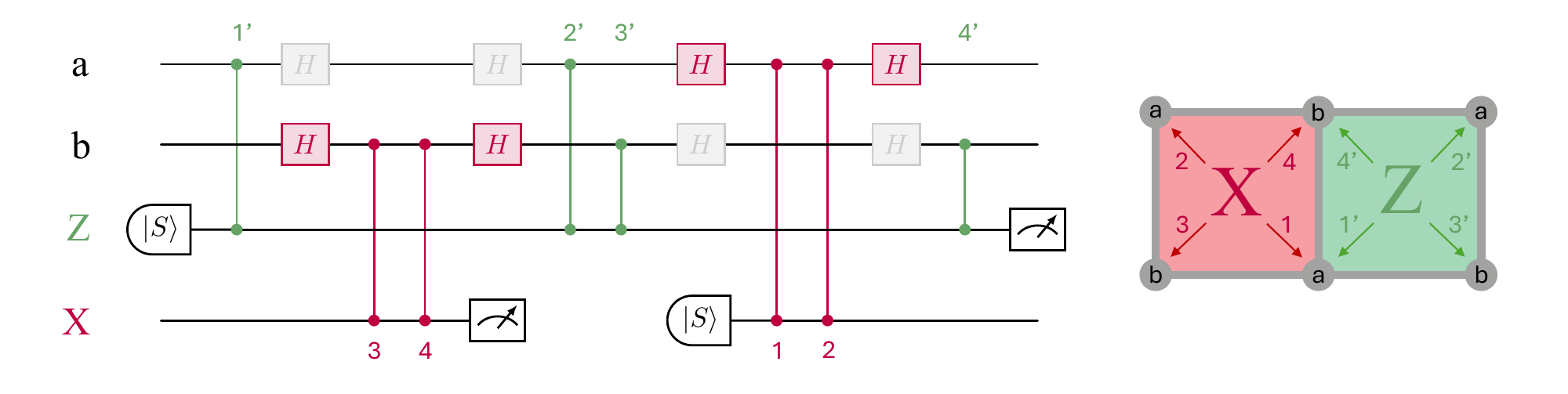}
    \caption{Circuit implementation of a cycle of $X$ and $Z$ stabilisers over one unit cell of the code. This implementation respects both the ordering avoiding hook errors, and makes use of semi-global Hadamard gates only. The a/b indexation for the data qubits is chosen so as to respect the periodicity of the lattice. The numbers (1 to 4 for $X$, 1' to 4' for $Z$) indicate the order in which gates should be implemented to avoid hook errors.}
    \label{fig:X_Z_interleaved}
\end{figure}

The two previous sections respectively addressed the questions of gate ordering and individual implementation of all types of stabilisers. The last step is now to give a protocol to combine these two elements and obtain the full circuit implementation of a stabiliser cycle. At a given time step, the qubits are scheduled to undergo one of the following operations: shuttling; initialisation or measurement; or single- or two-qubit gates. In the latter case, gates can be separated in two sets: gates used for $Z$ parity measurement (single $CZ$'s) (set 1) and gates used for $X$ parity measurement ($H$-$CZ$-$H$ sequence) (set 2). We include the $Y$ parity measurement of the twist in set 2 as the $S$ gates can be implemented locally and are thus not problematic. The idea is then simply to perform these two sets of gates one after the other, and the undesired Hadamards will simplify. Note that the same technique was already used in Fig. \ref{fig:circuit_all_stabs}. Additionally, by following the order `set 1 - set 2 - set 2 - set 1' over two consecutive rounds of stabiliser measurements, one can further halve the number of Hadamards. This protocol is illustrated in Fig. \ref{fig:X_Z_interleaved} in the case of no dislocations or twists (but the same idea would apply if they had to be measured too). Gate set 1 is represented in green and gate set 2 in red. 

\subsection{State injection} \label{sec:state_injection}

So far, we have shown how to implement all stabilisers involved in regular error correction as well as lattice surgery on the gate level. The final ingredient needed to enable universal quantum computation is state injection. The first step is to determine which logical states one would want to initialise.

The first one is the ancilla state used for multi-qubit measurements. Let us denote $P_i$ (resp. $P_{i,ancilla}$) the Pauli operators applied to the $i$-th logical data (resp. ancilla) qubit. For the logical measurement of $P_1 \otimes ... \otimes P_n$, Fig. 44 of \cite{Litinski_2019} prescribes the initialisation of the mediating logical ancilla in the $\ket{+}^{\otimes n-1}$ state, and measurement of the operators $Z_{i,ancilla} \otimes P_i$ via lattice surgery. In our case, as we can only prepare $\ket{0}$ states, preparing $\ket{+}$ states would require the application of an additional Hadamard. Therefore, it is simplest to initialise the logical ancilla in the $\ket{0}^{\otimes n-1}$ state and measure $X_{i,ancilla} \otimes P_i$ instead, which is strictly equivalent. Besides, as explained in \cite{Litinski_2019}, the ancilla is insensitive to $Z$ errors, as it is prepared in $Z$ eigenstates. This is particularly handy for us as our code is more prone to $Z$ errors as a result of shuttling (whose precise noise model will be studied in the next subsection).

The second logical state one would want to prepare is the magic state $\ket{m}=(\ket{0} + \mathrm{e}^{i\pi/4}\ket{1})/\sqrt{2}$.
To do so, we use a modified version of \cite{Li_2015, Lao2022}. The main obstacle in the direct implementation of these protocols on our device is the restriction on single-qubit gates. In particular, we cannot prepare a $\ket{+}$ state without applying a global Hadamard. Yet, these schemes require the preparation of the data qubits in both the $\ket{0}$ and $\ket{+}$ states. This issue can be circumvented by staggering the initialisation of the data qubits. More concretely, one would first prepare all qubits that should be initialised in the $\ket{+}$ state, as well as the physical magic state, in the $\ket{0}$ state. A global Hadamard is then applied, bringing all these qubits to the $\ket{+}$ state. Then a local phase gate can be applied on the physical magic state, thereby preparing it in the $\ket{m}$ state. All remaining qubits can then be prepared in the $\ket{0}$ state, and the state injection procedure of \cite{Li_2015, Lao2022} can be performed normally.

One important thing to note however is that whenever a global Hadamard is applied, it does not just target the patch of surface code one intends to prepare, but also all other data qubits in the device. As a result, a transversal $H$ gate is applied to all logical qubits in the middle of the quantum computation. This results in the application of logical Hadamards, as well as the permutation of the $X$ and $Z$ stabilisers of each surface code. While this could seem problematic, these additional Hadamards can easily be undone by the application of another logical Hadamard on all qubits. This is no different than any other Clifford gate in the circuit and can thus be performed virtually, by changing the basis of the following gates and measurements.

\subsection{Performance simulations under realistic noise}

In the previous sections, we showed that universal error correction is possible from a theoretical point of view on a 2$\times$N spin-qubit device with strong constraints, such as the access to semi-global single-qubit gates only (apart from phase gates). In the following, we argue that our proposition is also practical, \textit{i.e.} that the required physical error rates and resource requirements for such a device are not experimentally out of reach.

\subsubsection{Noise model}

First, let us discuss the noise model we adopt to estimate the performance of our system. The main ingredient we extensively make use of here and that requires modelling is shuttling.
When an electron is shuttled, its $g$-factor varies due to inhomogeneities in the device, which induces unwanted phase rotations in the laboratory reference frame. Systematic rotations can be calibrated away, so that we are effectively describing an electron in a shuttled reference frame \cite{Buonacorsi_2019, Buonacorsi_2020, Langrock_2023}.
When the phase is not exactly cancelled out owing to incorrect calibration, one can expect the data qubits of the device to each undergo some small coherent rotation (remember that only data qubits are shuttled). As coherent errors are difficult to simulate classically and potentially more harmful than stochastic errors, twirling is generally used to transform them into Pauli errors \cite{Wallman_2016}. However, this method is unfortunately not fully permitted in our device as some of the single-qubit gates can only be implemented semi-globally.
As such, the only natural candidate for a twirling gate set tailored to phase rotations and respecting our constraints is $W = \{I, X_1\otimes...\otimes X_n\}$ (with $I$ the identity gate, $X_i$ the Pauli $X$ gate on data qubit $i$ and $n$ the number of data qubits). Nonetheless, one can easily see that a Pauli $Z$ error on an even number of data qubits commutes with both gates of $W$ and will thus not be extracted from the coherent phase rotations by the twirling gates (another, more mathematical justification can be obtained by using Eq. 7 of \cite{Cai_2019}). Therefore, only odd-order errors (\textit{i.e.} Pauli $Z$ errors affecting an odd number of qubits) will correctly be twirled. If this poses some difficulties for the exact simulation of our system, it should however not be fatal in terms of performance, as the logical-level noise has been shown to behave similarly for coherent and random errors in large enough surface codes \cite{Bravyi_2018}.

Therefore, even though the twirling process is not perfect, we still approximate the remaining errors by some random dephasing of probability $p_{sh}$ for simulation purposes, which is correct up to first order (as second-order $Z$ errors are not correctly twirled). $p_{sh}$ here represents the probability that a given data qubit is affected by a Pauli $Z$ error somewhere within a given stabiliser cycle. To first order it is therefore the sum of four contributions, corresponding to the four shuttles that are carried out within a cycle.
Ignoring non-adiabatic effects for now, Eq. 4 of \cite{Langrock_2023} shows that the probability of a dephasing error $\delta\phi^2/2$ \textit{for one shuttle} follows:
\begin{equation}
    \delta\phi^2/2 = 2 \frac{l_cL_s}{(vT_2^*)^2}
\end{equation}
with $L_s$ the shuttling distance, $v$ the shuttling speed, $T_2^*$ the characteristic dephasing time experienced by stationary spins and $l_c$ the coherence length of the dephasing noise due to shuttling. The latter arises from imperfect calibration of the dynamic reference frame used to absorb coherent rotations happening when electrons are shuttled. These calibrations errors stem from uncontrolled fluctuations caused by slow nuclear dynamics due to dipolar interaction or low-frequency 1/$f$ charge noise affecting the spins' $g$-factors. Now summing the contributions of the four shuttles happening during a stabiliser cycle, one gets:
\begin{equation}
    p_{adia} = 2 \frac{l_c\times 4dl_{dd}}{(vT_2^*)^2}
\end{equation}
where $l_{dd}$ is the spacing between two data qubits. The total shuttling distance over a whole stabiliser cycle is roughly $4dl_{dd}$, as we need to shuttle by $2d$ increments and back (remember that we are interleaving logical data qubits and logical ancilla qubits, both of them of distance $d$, see Fig. \ref{fig:qubit_layout}).

Nonetheless, the above analysis would not be complete if we did not consider dominant non-adiabatic effects. The lowest energy splitting in silicon quantum dots is the valley splitting $E_{VS}$, which corresponds to the gap between the two out-of-plane conduction bands (or valleys). These two valley states are additionally characterised by distinct $g$-factors \cite{Cifuentes_2024}. Consequently, an electron in the excited valley state, as opposed to the expected ground valley state, will precess at an unknown frequency, leading to unwanted phase rotations. One can estimate the amount of non-adiabaticity as follows: $E_{VS}$ typically ranges between 10 and 200$\mu$eV for SiGe or can exceed 500$\mu$eV in SiMOS architectures \cite{Langrock_2023}. Furthermore, the two valleys are site-dependent, meaning that they depend on the position $x(t)$ of the electron in the device. Assuming that the length scale of the electron wavefunction is roughly $l_x=50$nm, one can assume that in the worst-case scenario the electron's valley state would switch every 50nm, leading to a characteristic energy $hv/l_x=0.83\mu$eV, where $h$ is Planck's constant. This is only 12 times smaller than 10$\mu$eV, \textit{i.e.} the worst-case value for $E_{VS}$, justifying the inclusion of these non-adiabatic effects in our modelling. Further modelling and simulations found in Appendix \ref{appendix:valley} lead to additional dephasing errors due to the valley state of $p_{dia}/4d=1.4\times10^{-6}$. This value will be used in all the simulations of the paper. The final shuttling-induced dephasing probability is then given by $p_{sh}=p_{adia}+p_{dia}$.

With these considerations in mind, we now have all necessary ingredients to simulate the performance of our device. Our full noise model thus includes $(i)$ dephasing channels of probability $p_{sh}=p_{adia}+p_{dia}$ on the data qubits at the beginning of each stabiliser cycle; $(ii)$ depolarising channels of error rate $p$ after each gate of the stabiliser circuit (including twirling gates and cancelled out Hadamards arising from their semi-global implementation); $(iii)$ depolarising channels of error rate $p$ after any initialisation; $(iv)$ classical flip of any measurement outcome with probability $p$. 

Note that here we ignore the idling errors --- these account for errors that idle qubits accumulate while others are being operated on. Indeed, spin-coherence times $T_2$ exceeding 20ms have been demonstrated for purified silicon electron spins \cite{Veldhorst_2014}, which is 4 orders of magnitude above initialisation, measurement and single- and two-qubit gate times (at most a few microseconds \cite{MAUNEB.M2012Csoi, Philips_2022, Zheng_2019, Takeda_2024}).
These $T_2$ times are obtained via dynamical decoupling, which can be implemented by periodically flipping all the spins in the device. In our case, it is simplest to flip all spins at the same time, as this is permitted by the semi-global pulse we are assuming.

In the following simulations, we fix the gate noise $p$ to $0.1\%$, so as to be below the usual circuit-level noise threshold of the surface code (around $0.7\%$ \cite{raussendorfTopologicalFaulttoleranceCluster2007}). This is consistent with experiments showing silicon spin qubit fidelities exceeding $99.9\%$ \cite{Yoneda_2017}. The dephasing time $T_2^*$ will range between $1\mu$s and $8\mu$s, which has been achieved \cite{Burkard_2023, Yoneda_2017}, and is well below demonstrated dephasing times of $100\mu$s \cite{Takeda_2024}.
For the estimation of $p_{sh}$, we fix $v=10$m/s, $l_c=100$nm and $l_{dd}=140$nm \cite{Langrock_2023}. This choice of 140nm for the distance between two data qubits takes into account additional space for the clavier gates required by the conveyor-belt mode of shuttling, and leaves extra space for singlet-triplet ancilla qubits and measurement apparatuses in the static row. These values in turn give a dephasing probability \textit{per shuttling increment} $p_{sh}/4d$ between $5.8\times 10^{-6}$ and $2.7\times 10^{-4}$.

The most optimistic end of this range corresponds to a more mature technology than the early demonstrations reported so far, which now approach $10^{-4}$ \cite{Yoneda_2021, Seidler_2022, desmet2024highfidelitysinglespinshuttlingsilicon}.
However the task of low-noise shuttling is receiving rapidly increasing attention, notably with recent proposals suggesting that $g$-factor disorder might help improve shuttling fidelities rather than hinder them \cite{Bosco_2024}. We will thus assume that these will keep improving.

As the surface code is a CSS code, $X$ and $Z$ errors can be analysed separately. Since our code is more prone to $Z$ type errors, we focus only on them.
Therefore, we numerically evaluate the $X$ logical error rate of a single distance-$d$ wide surface code used in memory mode via $N_\text{runs}$ runs of Monte Carlo simulations, with $N_\text{runs}$ ranging between 10,000 and 1,000,000 depending on the target logical error rates. In each run, random errors are injected in the code according to the error rates $p$ and $p_{sh}$, affecting both data and ancilla qubits for $d$ rounds of stabiliser measurements. The syndrome they create is then decoded via Minimum Weight Perfect Matching \cite{Fowler_2012}. The initial errors and the correction are then added to determine if the $X$ logical operator value was flipped.

\subsubsection{Simulations}

\begin{figure}
    \centering
    \includegraphics[width=\linewidth]{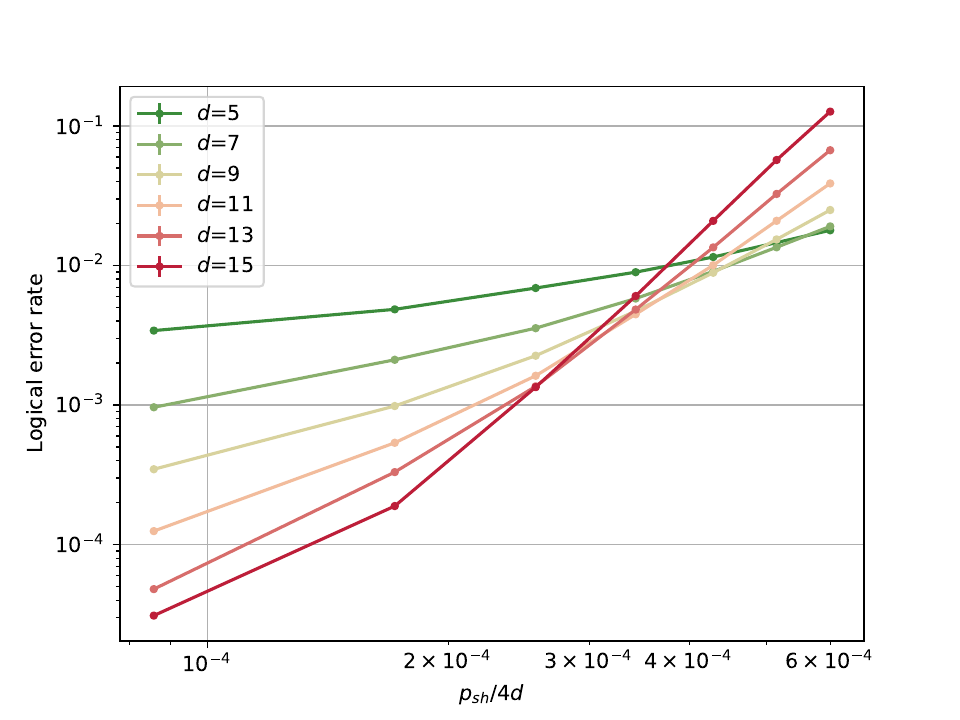}
    \caption{Logical error rate of a wide surface code against $p_{sh}/4d$, \textit{i.e.} the probability that one shuttling increment introduces a dephasing error, plotted for several code distance $d$. The gate, measurement and initialisation noise $p$ is set to $0.1\%$.}
    \label{fig:2xN_no_threshold}
\end{figure}

We first plot in Fig. \ref{fig:2xN_no_threshold} the logical error rate for several code distances $d$, against $p_{sh}/4d$ \textit{i.e.} the shuttling error per shuttling increment.
The main observation of the plot is that the lines corresponding to surface codes of various distances do not cross at a single point. This is expected as lines crossing at the same point, equivalent to the existence of a threshold at this crossing point, should only appear if the noise model respects certain properties. In particular, the noise strength should not scale with the code size \cite{Knill_1998}.
Yet, here we assumed that $p_{sh} \propto 4d$. As a result, the crossing points of subsequent-distance surface codes cross at lower and lower values of $p$. In other words, there does not seem to be any value of $p$ below which increasing the code distance consistently reduces the error rate. Rather, there is a constant trade-off between increasing the code distance for more powerful error correction, and keeping it low enough to keep the noise due to shuttling manageable.

\begin{figure}
    \centering
    \includegraphics[width=\linewidth]{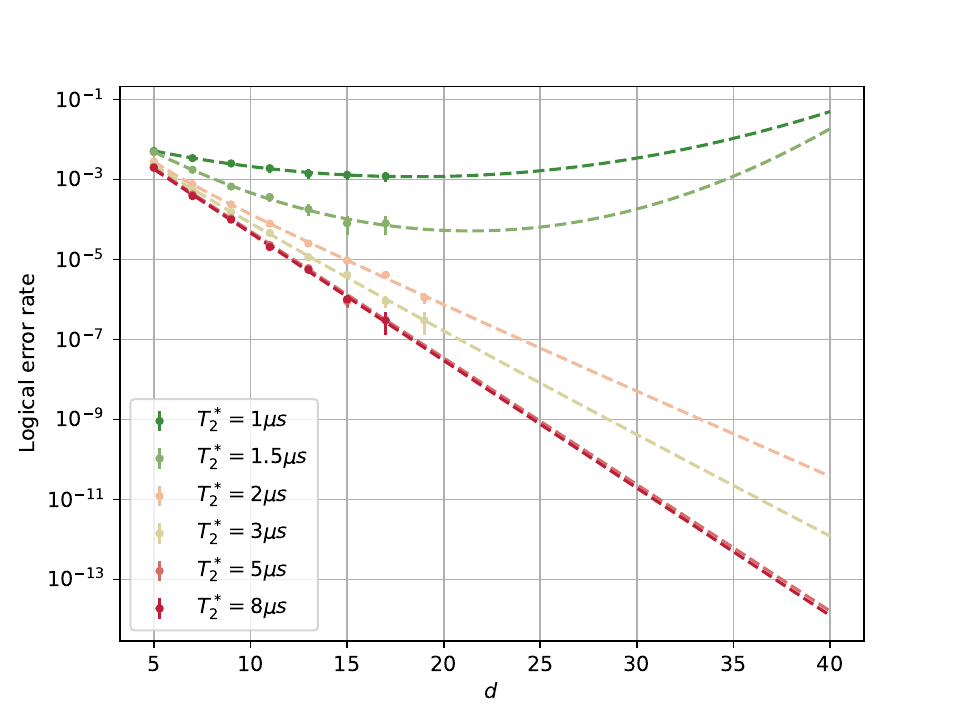}
    \caption{Logical error rate of a wide surface code against the code distance, plotted for several values of the stationary dephasing time $T_2^*$. The gate noise $p$ is set to $0.1\%$. The dots represent the simulation data, their error bars quantifying the sampling error. The dashed lines correspond to a fit of the data according to the trial function of Eq. \ref{eq:fitting_function}.}
    \label{fig:2xN_ 2xN_log_error}
\end{figure}

The anticipated non-existence of a threshold does not present any basic issue for our approach. Since the ideas presented here are intended for the early fault-tolerant era, the relevant quantity is always the logical error rate that can be achieved for realistic experimental parameters and specific system sizes.
In Fig. \ref{fig:2xN_ 2xN_log_error}, we thus plot the logical error rate against $d$ for $T_2^*$ ranging between $1\mu$s and $8\mu$s. The simulation data, represented by small dots with error bars (quantifying the sampling error), is then fitted with the following trial function:
\begin{equation} \label{eq:fitting_function}
    p_\mathrm{log}(d) = A (\alpha + \beta d)^{\gamma d+\delta}.
    \end{equation}
The motivation for this function is the following. If one denotes $p_{tot}$ the total noise experienced by the system, the surface code promises an exponential suppression of the errors of the form $p_\mathrm{log}(d) \propto (p_{tot}/p_{th})^{d/2}$ with $p_{th}$ a constant. Then, up to first order, $p_{tot}$ can be written as a sum of the gate noise $p$ and the shuttling noise $p_{sh}$, where the latter is proportional to the code distance $d$.
The consequence of $p_{tot}$'s dependence on $d$ is a sub-exponential suppression of errors, or worse, an increase in the error rate when the distance of the code becomes too large. 
Fortunately, with dephasing time that is large enough yet experimentally plausible, low enough error rates (around $10^{-13}$) can be reached before this phenomenon occurs.

\subsection{Resource estimation for universal quantum computation}

With all the above ingredients, we can now estimate the resources required for our device to achieve two meaningful milestones. In all the following, we set a spins' stationary dephasing time $T_2^*=8\mu$s, a per-dot valley noise $p_{dia}/4d=1.4\times10^{-6}$ and a gate noise $p=0.1\%$. The first task we focus on is estimating the size $N$ that would bring the 2$\times$N architecture beyond the NISQ regime. More concretely, we want to estimate $N$ such that the device would contain 50 logical qubits, able to interact through a universal set of gates, and with logical error rates of at most $0.01\%$ (ten times lower than physical error rates). Using Fig.~\ref{fig:2xN_ 2xN_log_error}, one can see that distance-9 wide surface codes are sufficient. Moreover, following our state injection protocol (Section \ref{sec:state_injection}), the probability of preparing a magic state in the wrong state is of the order of the physical error rate of the operations that created it, that is $p$ and $p_{sh}$. To first order, the probability $p_{\text{mag}}$ of initialising an erroneous logical magic state is thus of the order of $p+p_{sh}=0.103\%$ (ignoring small constant factors, see Eq. 1 of \cite{Lao2022}). This lowers to $35p_{\text{mag}}^3 = 4\times 10^{-8}$ after a 15-to-1 distillation protocol --- way below our target logical error rate of $0.01\%$. To meet our above requirements, one would therefore need $2\times (50 + 15) = 130$ distance-9 surface code patches (including logical data qubits and magic state distillation factory, and doubling this number to account for the logical ancilla bus connecting all these logical qubits, see Fig. \ref{fig:qubit_layout}). This translates into an overall size of $N=2\times 10^4$ physical data qubits.

Beyond simply outperforming NISQ on the paper, what one would actually want from a fault-tolerant quantum computer is to be able to run meaningful quantum algorithms. The second task we will thus focus on is estimating the ground-state energy of a 2D Hubbard model Hamiltonian. This can be solved by preparing the unitary $\mathcal{W}(H)=\mathrm{e}^{\mathrm{i \, arccos}(H)}$ by qubitisation, then passing it to phase estimation. An optimised protocol to run this algorithm can be found in \cite{Babbush_2018} along with the resource requirements for the smallest instance of this problem that is within the classically intractable regime, \textit{i.e.} a $6\times 6$ Hubbard model. For this grid size, 100 logical qubits and $10^8$ logical $T$ gates would be required.

Let us set a spins' stationary dephasing time $T_2^*=8\mu$s and a gate noise $p=0.1\%$. Let us assume that both the probability of a logical error impacting \textit{any} individual $T$ gate, and the probability of an error impacting \textit{any} of the logical qubits in the circuit, is below $10\%$. The algorithm thus fails with a probability of around $20\%$. The quantum computation can then be run multiple times in parallel to exponentially increase the success probability. Following the procedure of \cite{Litinski_2019}, one can estimate the physical resources needed to obtain the correct outcome.

\paragraph{Magic state distillation} The error rate of each individual $T$ gate must be under $10^{-9}$ to ensure that the failure probability of any of the $10^{8}$ gates is under $10\%$.
Again, the probability of initialising an incorrect logical magic state before distillation is $p_{\text{mag}}\sim p+p_{sh}$.
Assuming that the distance of our surface code is under 40 (which we will verify later on), we guarantee $p_{sh} < 0.02\%$. Adding the gate noise, we obtain a total noise $p_{\text{mag}}\lesssim 0.12\%$. Using the 116-to-12 distillation protocol, the probability of outputting incorrectly distilled magic states is $41.25p_{\text{mag}}^4 \lesssim 10^{-10}$, which is under our target per-gate error of $10^{-9}$.
44 surface codes patches are needed to run the 116-to-12 distillation protocol, which outputs 12 logical magic states in 99$d$ time steps with $89\%$ success probability \cite{Litinski_2019}. The initialisation time of one correct magic state is thus on average $9.27d$ time steps. The magic states can then be consumed one by one in the quantum computation, which requires $d$ time steps for each state. The overall time cost of the quantum computation (to prepare and consume $10^8$ magic states) is thus $10.27d\times 10^8$. It requires 100 logical qubits to store the logical information and 44 surface code patches in the distillation block (not counting the logical ancilla). Adding the long logical ancilla bus mediating all interactions, the logical qubit count doubles to 288.

\paragraph{Code distance} Given the total number of logical qubits and the computation time, in order to set the probability of a logical error affecting any of the logical qubits during the computation under $10\%$, one requires the per-logical-qubit error rate $p_L$ to be below
\begin{equation}
    p_L < 0.1 / (288 \times 10.27\times 10^8) = 3 \times 10^{-13}.
\end{equation}
By  using the fitting function of Fig. \ref{fig:2xN_ 2xN_log_error}, we can choose the code distance to be $d=36$. This is indeed within the upper bound of 40 that we set earlier.

\paragraph{Summary} Therefore, in order to run one instance of a $6\times 6$ Hubbard model (which is within the classically intractable regime) on our device, one would need 288 wide qubits of distance $d=36$, that is $1.4\times 10^6$ physical (data and ancilla) qubits. The algorithm would run in $10.27d\times 10^8$ stabiliser cycles.

The total shuttling time over a stabiliser cycle is $t_{sh}=4dl_{dd}/v$, where $l_{dd}=140$nm is the distance between two consecutive data qubits, and $v=10$m/s is the shuttling speed. Thus $t_{sh}=2\mu$s. Besides, while two-qubit gates can be implemented much faster (around $0.1\mu$s \cite{MAUNEB.M2012Csoi}), initialisation, measurement and single-qubit gates all take around a microsecond \cite{Philips_2022, Zheng_2019, Takeda_2024}. We can thus realistically assume an overall stabiliser cycle time of $6\mu$s, which would bring the computation time to a total of 2.5 days. While this time may seem relatively daunting, one must remember that we exclusively used experimental parameters that have been achieved, ignoring any further optimisation that will surely happen in the following years before this kind of device can be implemented.
Our shuttling noise model, informed by earlier theoretical studies (esp. Ref.\,\cite{Langrock_2023}), assumes levels of imperfection that are smaller than existing early demonstrations; we are confident that improving experimental techniques driven by rapidly increasing community interest will reach the domain that we have simulated. From Fig. \ref{fig:2xN_ 2xN_log_error}, one observes that for the more optimistic coherence times we have a scenario where shuttling noise is among the least significant factors in the model (as there is very little variation in the logical error rate when $T_2^*$ is decreased from $8\mu$s to $5\mu$s, \textit{i.e.} when the shuttling error per increment varies between $6\times 10^{-6}$ and $1.3\times 10^{-5}$). Further increasing $T_2^*$ to experimentally achieved values of $20\mu$s \cite{Struck_2020} would thus only have marginal effects on the performance here. Rather, a limiting factor seems to be the gate infidelity $p=0.1\%$. If this quantity could be improved by only a factor 2, logical error rates would drastically reduce due to almost exponentially scaling suppression (as long as the shuttling noise is kept low enough, \textit{e.g.} by enforcing coherence times of $20\mu$s this time).
Moreover, the assumed speed of the various operations corresponds to already-accomplished demonstrations, and does not begin to approach any fundamental physical limitations. One might reasonably expect orders of magnitude improvement over generations of such devices.

\section{Quantum memories with general qLDPC codes}

In section \ref{sec:framework} we showed that our device is suitable for the implementation of various classes of qLDPC codes beyond the surface code. These codes are known for their high performance and can fully exploit the connectivity (beyond that required by the surface code) which is provided by our shuttling-based system. Therefore, in this section, we explore the potential of such non-local codes, constructed by either allowing more ancilla-data qubit interactions, or by increasing the shuttling distance. Here, we only study their use as memory codes.

The gate ordering we choose in the stabiliser circuits is not optimised. All $X$ stabilisers will be implemented when the data qubits are shuttled one way, followed by all $Z$ stabilisers when the data qubits are shuttled back. Regarding the noise model, we adopt the same as in the previous section --- circuit-level noise plus additional dephasing due to shuttling (proportional to the shuttling distance). Due to the more complex data-ancilla interactions and to our non-optimal gate ordering, not all qubits perform their entangling operations synchronously (while it is the case for the surface code). Consequently, the data qubit rail might have to stop and start more often, and qubits that are not interacting will have to stay idle. To model this, we introduce an additional noise parameter $p_{idle}$, and add depolarising channels of probability $p_{idle}$ every time a qubit is idle while others are interacting. Note that we neglected this in the previous section due to the high coherence time of silicon spin qubits.

The numerical experiments are run under various values of the gate noise $p$, the idling noise $p_{idle}$ and the dephasing time $T_2^*$. The per-dot valley noise is still fixed to $p_{dia}/4d=1.4\times10^{-6}$. Each code is simulated for $d$ rounds of stabiliser cycles, where $d$ is the code's estimated distance. The decoding is performed via \textit{min-sum} BP-OSD, a limit of 32 iterations, a scaling factor of 0.625 and a serial schedule \cite{Panteleev_2021}. The logical error rate \textit{per round per logical qubit} $p_{log}$ defined below is then plotted and used to compare the performance of various codes \cite{bravyi2023highthreshold}:
\begin{equation}
    p_{log} = 1 - (1 - P_L(k,d))^{1/kd}.
\end{equation}
Here, $P_L(k,d)$ is the probability that a logical error happens on at least one of the $k$ logical qubits of an $[n,k,d]$ code running for $d$ rounds of stabiliser cycles.

\subsection{Hypergraph product code with a repetition code} \label{sec:hgp}

As explained in Section \ref{sec:code_examples}, one class of codes that is particularly suitable for our architecture are the hypergraph product (HGP) codes constructed with the repetition code as one of the seed codes. Indeed, let us suppose that shuttling is performed in the vertical direction, and that a HGP code is built from a classical LDPC code placed in the direction of shuttling, and a repetition code placed in the orthogonal direction. Let us first assume that both codes have distance $d$. It follows that horizontal interactions have length at most 1 in the 2D grid (length of the repetition code's interactions), which translates into length $d$ in the linearised 2$\times$N architecture. Therefore, any ancilla qubit has to interact with data qubits that are at most $d$ increments away, and compared to the surface code, the shuttling distance is not increased. Rather, the only difference with the latter is that the data qubit row has to stop more often, as stabilisers may have higher weights, but also as it may not be possible to synchronise the interactions as much as for the surface code (adopting the framework of Section \ref{sec:framework}, the code's check matrix has more non-zero diagonals). Therefore, implementing such HGP code comes at the cost of increased idling noise (but not increased shuttling noise).
This additional noise can be compensated for if a \textit{good} classical code is chosen for the product with the repetition code, as a higher-rate quantum code will be generated, thereby reducing the qubit overhead.

In the following, we study the example of a [234,3,8] HGP code. The code is obtained by taking the product of a classical [17,3,8] LDPC code $H_1$ (generated randomly until good parameters were obtained, see Appendix \ref{appendix:hgp_matrix}) and a distance-8 repetition code $H_2$. To limit the number of two-qubit gates, which are expected to be the leading source of noise, we generate $H_1$ by enforcing two 1's per row on average, thereby guaranteeing that the stabilisers of the quantum code are weight-four on average. This code is more compact than the surface code as it encodes $k=3$ logical qubits with $(17+14)\times (8+7)=465$ physical qubits (data and ancilla). On the contrary, one would require $k\times (2d-1)^2 = 675$ physical qubits to encode the same amount of logical information with same-distance surface codes.

\begin{figure}
    \centering
    \includegraphics[width=\linewidth]{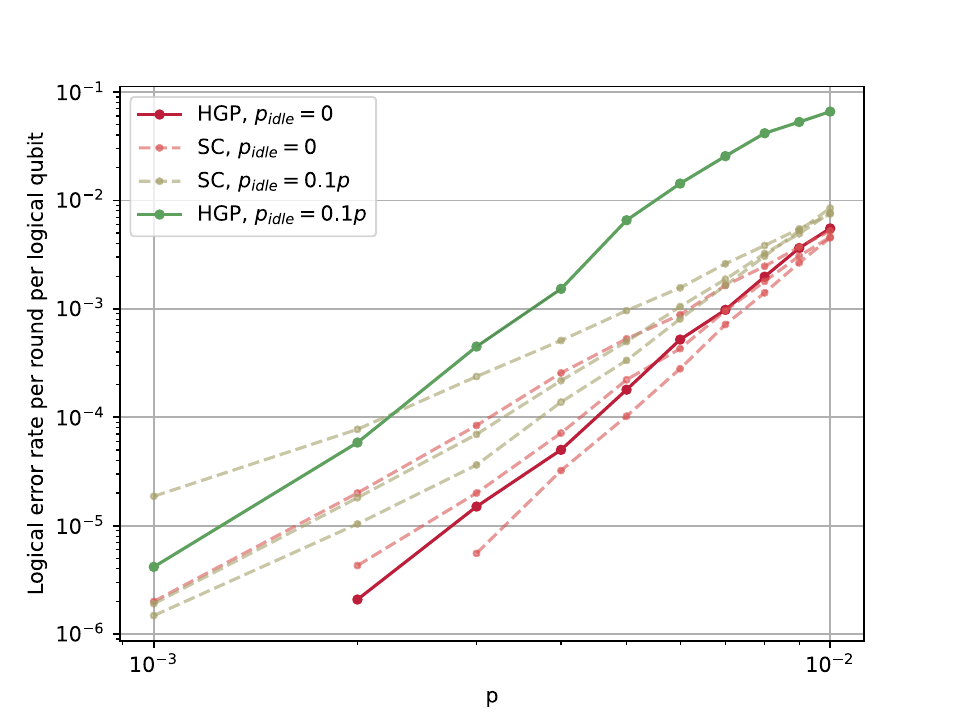}
    \caption{Logical error rate per round per logical qubit of a [234,3,8] hypergraph product code (HGP, solid lines) and surface codes of various distances $d$ (SC, dashed lines), against the circuit-level noise parameter $p$. The HGP code is built from the product of a repetition code and a classical code generated randomly. The dephasing time $T_2^*$ is set to $8\mu$s, and the logical performance of the HGP and surface codes are plotted for $p_{idle}=0$ and $0.1p$. For both levels of idling noise, the three surface code lines correspond from top to bottom to a distance of 5, 7 and 9 respectively.}
    \label{fig:hgp}
\end{figure}

Fig. \ref{fig:hgp} compares the logical error rates per round per logical qubit $p_{log}$ of the [234,3,8] HGP code and of surface codes of various distances. The shuttling noise is fixed by setting $T_2^*=8\mu$s. As explained above, the HGP code experiences the same amount of shuttling error as the surface code, but higher idling noise. Indeed, its two-qubit gates cannot be implemented synchronously due to the randomness of the matrix $H_1$ (it contains many non-zero diagonals). Qubits that are not interacting thus have to stay idle, which induces depolarising errors at a rate $p_{idle}$. We thus plotted $p_{log}$ for two levels of idling errors: $p_{idle}=0$ and $p_{idle}=0.1p$. One can observe that for no idling errors and for a relevant range of gate noise $p$, the HGP code performs similarly as surface codes with comparable distances, yet requiring $30\%$ less qubits. As expected however, when the idling noise is increased, the HGP code's performance falls dramatically while the surface code's is very moderately affected. 

While observing this, one needs to take into account that we did not optimise the stabiliser circuit. An interleaved scheme of $X$ and $Z$ measurements could drastically reduce the overall idling noise.
Besides, in the specific case of silicon spin qubits, $p_{idle}$ is expected to be extremely small. Indeed, given the experimental values observed for the coherence time $T_2$, around 20ms \cite{Veldhorst_2014}, and the slower gates times, around $T_{gate}=1\mu$s \cite{Philips_2022}, one would obtain an idling error probability of
\begin{equation*}
    p_{idle} = 1-\mathrm{e}^{-T_{gate}/T_2} = 5\times 10^{-5}.
\end{equation*}
When using this value, one would obtain the same qualitative performance as for the $p_{idle}=0$ case.

Furthermore, here we set a single parameter $p_{idle}$ quantifying the idling error. It would be more comprehensive to distinguish different $p_{idle}$'s depending on the operation that is being waited for. Specifically, as two-qubit gates are one order of magnitude faster than single-qubit gates, they would be responsible for shorter wait times, thus lower idling errors. Indeed, it so happens that the difference between the HGP code and the surface code lies in the additional idling times due to asynchronous two-qubit gates, which are the fastest operations in the circuit.

Lastly, do note that the crossing point of the surface code lines around $1\%$ is not a threshold in the common sense, as we are here plotting the logical error rate per round per logical qubit.

\subsection{Generalised bicycle codes}

In the previous section, we studied a type of code that did not require an increase of the shuttling distance compared to the surface code, but we did not set any constraint on the number of stops along the way. We therefore guaranteed both a shuttling and idling noise scaling as $O(\sqrt{n})$, where $n$ is the number of data qubits in the code block. Here we study another paradigm, where we do not restrict the shuttling distance, but instead bound the number of stops: $p_{sh}=O(n)$ and $p_{idle}=O(1)$. Codes that satisfy this are generalised bicycle codes \cite{Panteleev_2021}, as their check matrices by definition contain a constant number of non-zero diagonals (see Section \ref{sec:code_examples}).

In the following, we study the performance of a [126,28,8] generalised bicycle code (code A2 of \cite{Panteleev_2021}). This code contains contains 252 physical qubits (data and ancilla). Using the same number of physical qubits to encode $k=28$ logical qubits, one would need to use 28 surface codes of distance
\begin{equation*}
    d_{sc} = \frac{\sqrt{252/k}+1}{2} = 2.
\end{equation*}
This very small number stems from the relatively high rate of the code. However, tackling larger distances would require simulating much bigger codes which are beyond our numerical capabilities.

\begin{figure}
    \centering
    \includegraphics[width=\linewidth]{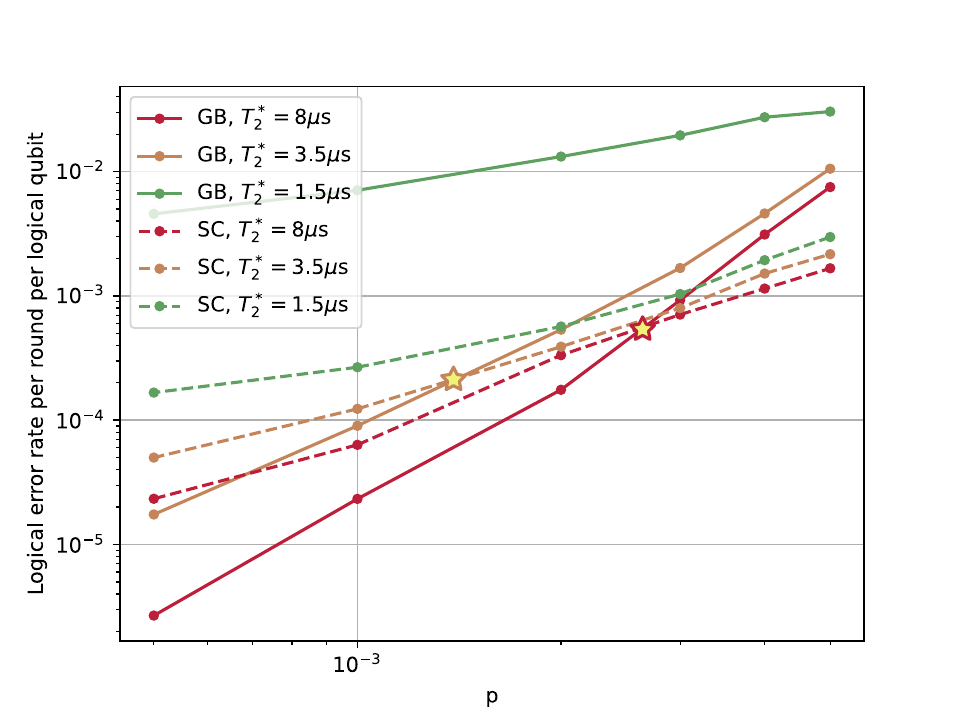}
    \caption{Logical error rate per round per logical qubit of a [126,28,8] generalised bicycle code \cite{Panteleev_2021} (GB, solid lines) and a distance-3 surface code (SC, dashed lines), against the circuit-level noise parameter $p$. The logical performance of both codes is plotted for $p_{idle}=0.01p$ and different levels of shuttling noise, controlled by the dephasing time $T_2^*$. For $T_2^*=3.5\mu$s and $8\mu$s, the noise level where generalised bicycle codes start to beat surface codes is indicated by a star.}
    \label{fig:generalised_bicycle}
\end{figure}

Similarly to the previous subsection, we plot in Fig.~\ref{fig:generalised_bicycle} the logical error rate per round per logical qubit of the [126,28,8] generalised bicycle code and compare it to a distance-3 surface code (the smallest surface code that can correct, not just detect, errors). We set $p_{idle}=0.01p$ (with $p$ the measurement, initialisation and gate noise) and plot the logical performance of the codes for different amounts of shuttling noise, which are parameterised by the static dephasing time $T_2^*$. Compared to the previous subsection idling noise does not dramatically impact the bicycle code, as it is formed from matrices with only $5$ non-zero diagonals. However, the code is more prone to shuttling errors as the shuttling distance is now of the order of the number of qubits $n$ (while it is of the order $d=\sqrt{n}$ for the surface code).

One can see that for low enough (yet experimentally achievable) shuttling noise, the [126,28,8] code outperforms surface codes with similar qubit overhead for gate noise as high as $0.3\%$, which is also within reach. If the gate noise can be further reduced to $0.05\%$, the general bicycle code would provide an advantage of around one order of magnitude compared to a distance-3 surface code (which also uses more qubits than the generalised bicycle code, as we were meant to compare with distance 2). Nevertheless, the performance of the generalised bicycle code is as expected more sensitive to the shuttling noise than the surface code, as it involves longer shuttles. But for the considered code size, the effect only becomes predominant for relatively low dephasing times. Of course, when considering larger codes, the demands on the dephasing time would have to be adjusted accordingly, to compensate for the even longer shuttles.

\subsection{Comparative performance} \label{sec:Comparative performance}

We here confirm the observations of this section via additional simulations aiming at further understanding the trade-off between resource requirements and error correction performance. Specifically, we examine the performance of the three previously considered code families under three different noise regimes: $(A)$ low idling noise but high shuttling noise; $(B)$ low shuttling noise but high idling noise and $(C)$ high idling noise and high shuttling noise. The results of such simulations are presented in Table \ref{tab:sc_hgp_gb_comparison}, where the theoretical parameters of the code are given, along with the logical error rate per round per logical qubit in the three noise cases. We aim to compare codes with similar overhead, defined as the total number of physical qubit per encoded logical qubit \textit{i.e.} $(n+n_{anc})/k$. In the first half of the table, we compare low-overhead codes: a distance-3 surface code (SC3), the [126,28,8] generalised bicycle code considered before (GB8) and a [40,3,4] hypergraph product code (HGP4). Its check matrix is given in Appendix \ref{appendix:hgp_matrix}. In the second half of the table, we focus on higher-overhead codes: a distance-7 surface code (SC7) and the [234,3,8] hypergraph product code considered before (HGP8). Note that in this case, we could not simulate \textit{good} generalised bicycle codes due to limited computational power. In each of the last three columns, the lowest error rate is highlighted, confirming the conclusions of the section: each code family considered in this paper excels in distinct noise regimes. Specifically, GB codes perform best when the shuttling noise is kept low; HGP codes do when the idling noise is low; and the SC wins when neither of them is sufficiently small. In the earliest fault-tolerant regimes, the latter scenario might be the most relevant, which justifies the use of the surface code, for which we also proved full computational power in the previous section.

\begin{table}[]
    \centering
    \begin{tabular}{ |c|c|>{\centering\arraybackslash}p{0.3cm}|c|c|c|c|c| }
     \cline{2-8}
    \multicolumn{1}{c|}{} & $\frac{n+n_{anc}}{k}$ & $d$ & \makecell{Shuttling\\noise\\scaling} & \makecell{Idling\\noise\\scaling} & $A$ & $B$ & $C$ \\
     \hline
     SC3  & 25 & 3 & $O(\sqrt{n})$ & $O(1)$ & $2.10^{-4}$ & $1.10^{-4}$ & $\mathbf{3.10^{-4}}$ \rule[1ex]{0pt}{1.7ex} \\ 
     GB  & 9 & 8 & $O(n)$ & $O(1)$ & $7.10^{-3}$ & $\mathbf{5.10^{-5}}$ & $7.10^{-3}$ \rule[1ex]{0pt}{1.7ex} \\ 
     HGP4 & 26 & 4 & $O(\sqrt{n})$ & $O(\sqrt{n})$ & $\mathbf{4.10^{-5}}$ & $1.10^{-3}$ & $1.10^{-3}$ \rule[1ex]{0pt}{1.7ex} \\ 
     \hline
     SC7  & 169 & 7 & $O(\sqrt{n})$ & $O(1)$ & $1.10^{-5}$ & $\mathbf{3.10^{-6}}$ & $\mathbf{1.10^{-5}}$ \rule[1ex]{0pt}{1.7ex} \\ 
     HGP8 & 155 & 8 & $O(\sqrt{n})$ & $O(\sqrt{n})$ & $\mathbf{8.10^{-6}}$ & $1.10^{-5}$ & $2.10^{-5}$ \rule[1ex]{0pt}{1.7ex} \\ 
     \hline
    \end{tabular}
    \caption{Comparative performance of three code families: surface code (SC), hypergraph product code (HGP) and generalised bicycle code (GB). See the main text for the precise definition of each code. The first three and last two rows respectively correspond to lower- and higher-overhead codes, where the overhead is defined as the number of physical qubits per logical qubits. The second column contains said overhead. The third column is the code distance. The fourth and fifth columns contain the shuttling and idling noise scalings of each code. Finally, in the last three columns the logical error rate per round per logical qubit is given for three different noise regimes: $(A)$ $T_2^*=1.5\mu$s, $p_{idle}=0$; $(B)$ $T_2^*=5\mu$s, $p_{idle}=0.1p$; and $(C)$ $T_2^*=1.5\mu$s, $p_{idle}=0.1p$. In each of these columns, the highest-performance low-overhead code, as well as the highest-performance high-overhead code is highlighted with bold letters.}
    \label{tab:sc_hgp_gb_comparison}
\end{table}

\section{Discussion}

We have investigated the feasibility of quantum error correction in a highly constrained experimental setup, specifically a 2$\times$N array of qubits where long-range interactions are enabled by shuttling one of rows of the array. By establishing a precise framework for determining code classes that naturally embed in our device, we showed that its strong constraints should theoretically not be an obstacle to early fault tolerance. This observation was then complemented by the design of an entire protocol permitting full universal quantum computation with the surface code. 

To show the practicality of our approach, we further tailored our protocol to silicon spin qubits, respecting their additional specificities and constraints, such as the difficulty to implement local Hadamard gates. This choice of platform was motivated by the high-fidelity shuttling capabilities that have been demonstrated \cite{Yoneda_2021, Seidler_2022, desmet2024highfidelitysinglespinshuttlingsilicon}, making it an ideal candidate for the implementation of our scheme. We then confirmed these theoretical protocols with extensive numerical simulations, showing that error rates as low as $10^{-13}$ can be reached with experimental parameters that have been achieved today, provided currently observed shuttling errors can be further suppressed to match our theoretical models. While this entails long quantum computations for running algorithms in the classically intractable regime, we are confident that the gate and shuttling performance will further improve in the coming years, making our proposal even more practical.

Furthermore, our exploration extends to the application of our device to more intricate qLDPC codes. Although more powerful than the surface code, their implementation comes at the cost of increased noise (either from longer shuttles or higher idling times). By simulating these codes and evaluating their performance in our constrained setup, we observed that this trade-off is in favour of the complex qLDPC codes when error rates are low enough (while still practically achievable). This underscores the versatility of our platform, even within the limitations of the 2$\times$N qubit array.

As future work, it would be interesting to engineer a slightly more complex architecture design that would let one include both the surface and better qLDPC codes in the same device and interface them. It would indeed prove advantageous to make use of the latter's strong error correction capabilities when used as memory codes. Whenever a logical qubit is idle for a significant amount of time during a quantum computation, one could take its logical information and store it in memory, before taking it out again when needed. If this store in-and-out scheme can theoretically be implemented via a modified lattice surgery protocol between a surface code and another type of qLDPC code \cite{Xu_2023}, one would still need to understand how to efficiently embed this in our device.

More generally, throughout this paper we have chosen to work within quite severe limitations, \textit{i.e.} the low-dimensional array and the restriction to semi-global Hadamard operations. We have established that even within these constraints computation is possible. Nonetheless, without further improvement, scaling up the current device to larger instances may prove very challenging. We identify two reasons for this. First, even for quantum algorithms slightly beyond the classically intractable regime, the run-times reach several days. This is partially due to our sub-optimal choice of logical qubit layout, where the logical information is stored in the first row of surface codes while the second row is only used as a logical ancilla bus (Fig. \ref{fig:qubit_layout}). While enabling all-to-all connectivity, this structure also introduces connectivity lockups that slow down the computation: for instance, when the leftmost and rightmost logical qubits are interacting through the ancilla bus, no other interaction can take place in parallel. On top of this, the second obstacle to large-scale computation in our current proposal is the sensitivity of the device to malfunctioning qubits. If a dot becomes inoperable, \textit{e.g.} due to a fabrication defect, and qubits cannot shuttle through it anymore, the array will be cut in two halves that cannot communicate anymore. This would likely lead to a failure of the algorithm.

To circumvent both the aforementioned issues, we envisage an extension of the strictly 2$\times$N qubit array presented in this paper: a lattice or web-like structure where long 2$\times$N filaments meet at occasional three- or four-way junctions through which qubits can shuttle. Such junctions could be well-spaced, so that the additional device complexity associated with each junction need not overlap with others. In addition to avoiding the problem of a single point of failure, such a geometry would no doubt afford a rich space of possibilities for novel codes and compilation strategies, thereby reducing run-times. They would additionally offer superior opportunities to interface the aforementioned better qLDPC codes with surface codes. This is an intriguing direction for future investigation.

\section*{Acknowledgments}
The authors are grateful to numerous supportive colleagues and especially Simon Benjamin and Fernando Gonz\'alez Zalba. They would also like to acknowledge the use of the University of Oxford Advanced Research Computing (ARC) facility \cite{richards_2015_22558} in carrying out this work and specifically the facilities made available from the EPSRC QCS Hub grant (agreement No. EP/T001062/1). The authors also acknowledge support from EPSRC's Robust and Reliable Quantum Computing (RoaRQ) project (EP/W032635/1), and EPSRC's Software Enabling Early Quantum Advantage (SEEQA) project (EP/Y004310/1). This research has also received funding from the European Union's Horizon 2020 Research and Innovation Programme under grant agreements No 951852.

\bibliography{ref}

\appendix

\section{Surface code's stabiliser circuit gate ordering} \label{appendix:gate_ordering}

Here we delve into the details of finding an optimal gate ordering for surface-code error correction on the 2$\times$N architecture. Indeed, the usual pattern that is used for the regular rotated surface code with N- and Z-shaped orderings \cite{Tomita_2014} leads to unnecessarily long shuttles. To see this, assume that the shuttling direction is vertical --- two consecutive data qubits within the same column (resp. row) of the surface code are thus separated by 1 (resp. $d$) shuttling increment(s). Therefore, a Z-shaped ordering would require to shuttle four times along rows, and it would yield a total shuttling distance of roughly $4d$. Our aim is to find an ordering that lets us implement all gates of the stabiliser cycle with a total shuttling distance of roughly $2d$ (which corresponds to having the data qubits do one round trip, not two).

\begin{figure}
    \centering
    \includegraphics[width=\linewidth]{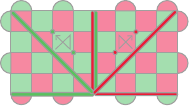}
    \caption{Representation of the wide surface code. $X$ and $Z$ stabilisers are respectively represented with red and green squares or half-disks. Several shortest-length representatives of the logical $X$ and $Z$ operators are drawn with horizontal, vertical and diagonal lines. The gray arrows represent an order in which $X$ and $Z$ stabilisers can be measured to avoid hook errors reducing the distance of the code. Examples of $X$ and $Z$ hook errors are respectively drawn with red and green disks: as they do not coincide with the logical operators, they do not reduce the distance of the code. The same applies to the regular rotated surface code.}
    \label{fig:wide_qubit}
\end{figure}

Furthermore, note that N- and Z- shaped orderings mentioned above do lead to distance-reducing hook errors in the wide surface code used in Section \ref{sec:universal_quantum_computation}. Indeed, its $X$ and $Z$ logical operators both have a horizontal and a vertical representative. Instead, one can measure the stabilisers as shown in Fig. \ref{fig:wide_qubit}. The shortest $Z$ logical operators are horizontal, vertical or diagonal. The latter type passes through the centres of $X$ stabilisers (red squares). However, using the measurement schedule represented by the gray arrow, $Z$ hook errors are on either diagonal of the $Z$ stabilisers (green squares), thus do not reduce the code distance. The same applies to $X$ hook errors. Besides, with the same reasoning, one can prove that the regular rotated surface code is protected just as well from hook errors when this ordering is used. In summary, whether it is for the regular (Fig. \ref{fig:RotSurface}) or wide (Fig. \ref{fig:wide_qubit}) surface code, this cross-like sequence is the one we consider.

Now, one could theoretically implement it by measuring the $X$ and $Z$ stabilisers separately on alternating rounds. However, this would unnecessarily increase the circuit depth and leave many qubits idle. Instead, a common solution is to interleave the gates of both stabilisers. This is possible provided the gates can be correctly commuted through each other so as to leave two neighbouring ancilla qubits disentangled (condition \ref{A1}); as well as gates can be implemented synchronously respecting the device layout and global shuttling of the data qubits (condition \ref{A2}).

\begin{figure}
    \centering
    \includegraphics[width=\linewidth]{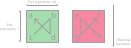}
    \caption{Notations for conditions (\ref{A1}), (\ref{A2}) and (\ref{A3}).}
    \label{fig:gate_ordering_conditions}
\end{figure}

Let us explain this more formally, and call time step $k$ the interval between shuttles $k-1$ and $k$. At a given time step, some entangling gates must be implemented between certain ancilla-data qubit pairs. Let us use the notations of Fig. \ref{fig:gate_ordering_conditions}, where each letter represents the time step when the ancilla and corresponding data qubit must be entangled. 

Condition (\ref{A1}) is verified if and only if \cite{Litinski_2018}
\begin{equation} \label{A1}
    ((b<e) \land (d<g)) \lor ((b>e) \land (d>g)).
\end{equation}
As for condition (\ref{A2}), it reads
\begin{equation} \label{A2}
    (a \equiv e[s]) \land (b \equiv f[s]) \land (c \equiv g[s]) \land (d \equiv h[s]).
\end{equation}
where $s$ is the number of shuttles required to go back to the data qubits' initial position.
This is because the device is laid out periodically and data qubits move as a whole along their shuttling track. This means that at any time step, ancilla qubits all face either their North-West, or North-East, or South-West, or South-East data qubit.

One last condition (\ref{A3}) can be added, enforcing the no-distance-reducing-hook-error ordering, which mathematically reads as
\begin{equation} \label{A3}
    (c<b<d<a) \land (h<e<g<f).
\end{equation}

Values for the time steps $a$ to $f$ respecting all three conditions are given in Fig. \ref{fig:gate_ordering} (here $s=4$). While their gates are indeed interleaved, $X$ and $Z$ stabilisers are nonetheless operated on a staggered fashion. The whole operation sequence, including gates, measurements, initialisations and shuttles is the following:
\begin{enumerate}
	\item entangle all ancilla qubits with their South-West data qubit
	\item shuttle by $d-1$ increments forwards
	\item entangle all ancilla qubits with their North-East data qubit; measure and reinitialise $X$ ancilla qubits
	\item shuttle by $1$ increment forwards
	\item entangle all ancilla qubits with their South-East data qubit
	\item shuttle by $d+1$ increments backwards
	\item entangle all ancilla qubits with their North-West data qubit; measure and reinitialise $Z$ ancilla qubits
	\item shuttle by $1$ increment forwards
	\item repeat
\end{enumerate}
Note that the shuttling increments given here do not include the additional shuttling accommodating the ancilla bus of Region B in Fig. \ref{fig:qubit_layout} ($n=1$). Besides, for the first round of stabiliser measurements, $X$ ancilla qubits should only start to undergo their entangling gates from step 3. Similarly, for the last round, only $X$ stabilisers should follow step 1 and 2.

With this protocol, in order to perform $N_r$ rounds of $X$ and $Z$ stabilisers measurements, one thus needs $4N_r+1$ shuttles and a total shuttling distance of $N_r(2d+2)+d-1$ (not including the logical ancilla bus of Fig. \ref{fig:qubit_layout}). One can also easily see that $4N_r$ and $N_r(2d+2)$ are the respective lower bounds for the number of shuttles and total shuttling distance, no matter what gate ordering is chosen. Indeed, a given four-body stabiliser trivially requires four steps to entangle the ancilla with all the data qubits, thus four shuttles. Moreover, its North-West and South-East data qubits are separated by a distance $d+1$, hence going back and forth between them requires a shuttling distance of $2d+2$. Therefore, apart from a small correction arising at the last round due to the staggered implementation of the $X$ and $Z$ stabilisers, our solution is optimal.

\section{Modelling for the valley degree of freedom} \label{appendix:valley}

The band structure of bulk silicon features six degenerate conduction bands: four in-plane and two out-of-plane. In the presence of gates in SiMOS or Si heterostructures, the six-fold degeneracy is lifted and the two out-of-plane bands become the lowest of the six: they are called the ground and excited valley states. Ideally always sitting in the ground valley state, an electron that is shuttled fast enough can non-adiabatically populate the excited valley state. As these states have been shown to exhibit distinct $g$-factors \cite{Cifuentes_2024}, the electron could then start to precess in an uncontrolled manner, causing unwanted phase rotations. This justifies the importance to control and estimate the impact of the excited valley state occupation.
To model this, we will use an extension of the valley state modelling in \cite{Langrock_2023}. 

Let us first introduce the position-dependent valley phase $\varphi_{VS}(x)$ and the bare valley splitting $E_{VS,0}$ (which models the valley splitting in the absence of the perturbations described below). The local two-level valley Hamiltonian can most generally be expressed as:
\begin{equation}
H_{loc}(x) = \frac{E_{VS,0}}{2} \left(\cos(\varphi_{VS}(x))\tau_x + \sin(\varphi_{VS}(x))\tau_y\right)
\end{equation}
where $\tau_x$ and $\tau_y$ are Pauli operators in the valley subspace.

Assuming low coupling of the spatial and valley degrees of freedom, an electron of spatial probability distribution $\rho(x)$ would thus experience an average valley Hamiltonian:
\begin{equation}
H_v(x_0) = \int \rho(x-x_0) H_{loc}(x_0)\mathrm{d}x
\end{equation}
Here we suppose that $\rho$ is a Gaussian of standard deviation $l_x$. Two extreme cases can be considered for $\varphi_{VS}$: smoothly varying or abruptly changing owing to the presence of atomic steps. In the $v=10$m/s regime, Fig. 1 of \cite{Langrock_2023} shows that the smooth interface model leads to higher noise: this is thus the model we will adopt.

That paper focuses on the simplest case of a linear gradient model: $\varphi_{VS}(x)=a_x x$. We slightly refine it by assuming that the gradient is not constant but instead slowly varying at the scale of the electron wavefunction. This means that we can adopt all equations derived in \cite{Langrock_2023}, while assuming a slow variation of $a_x$ to take disorder into account. In the instantaneous valley state basis, the final valley+orbital Hamiltonian is thus the following:
\begin{equation}
\tilde{H_v}(x_0) = \left(\frac{E_{VS}(x)}{2} \tau_z + \frac{\dot{\varphi}_{VS}(x)}{2} \tau_x\right) \otimes I + \frac{\Delta g(x)}{2} \frac{I-\tau_z}{2}  \otimes \sigma_z
\end{equation}
with:
\begin{align}
    E_{VS}(x) &= E_{VS,0}~\mathrm{exp}\left(-\frac{a_x(x)^2l_x^2}{4}\right) \\
    \dot{\varphi}_{VS}(x) &= \dot{a}_x(x) x + a_x(x) v(x) \label{eq:phi_dot}
\end{align}
$\sigma_z$ is the Pauli operator characterising the spin, $(I-\tau_z)/2$ is the projector onto the excited valley state and $\Delta g(x)$ models the difference in $g$-factor between the ground and excited valley states. We assume that the electron is smoothly shuttled back and forth along $2d$ dots separated by a distance $l_{dd}$ (to account for the ancilla bus of Fig. \ref{fig:qubit_layout}), such that
\begin{equation}
v(x)= \begin{cases}
  v, & \text{if}\ x < 2dl_{dd} \\
  -v, & \text{otherwise}
\end{cases}
\end{equation}
and
\begin{equation}
x(t)= \begin{cases}
  vt, & \text{if}\ t < 2dl_{dd}/v \\
  -vt + 4dl_{dd}, & \text{if}\ 2dl_{dd}/v \leq t < 4dl_{dd}/v
\end{cases}
\end{equation}
To model the disorder, we describe $a_x(x)$ with a smooth random walk of the form:
\begin{equation}
    a_x(x) = \sum_{k=1}^n \alpha_k \sin(\lambda_k x)
\end{equation}
with $n=20$ and $\lambda_k$ chosen randomly between $l_x$ and the maximum shuttled distance $2dl_{dd}$. These bounds guarantee a slow variation of the valley parameters on the scale of the electron wavefunction but over the whole landscape explored via shuttling. In the worst-case scenario, the ground and excited valley states should swap every $l_x$, meaning a variation of the valley phase of $\pi$. Thus a worst-case value for $a_x$ is $\pi/l_x$. For an $n$-step random walk with unitary steps ($\alpha_k=1$), the average maximum distance is $\langle \max_x a_x \rangle = \sqrt{\frac{n\pi}{2}}$. We therefore randomly sample $\alpha_k$ between 0 and 
\begin{equation}
    \alpha_{\text{max}} = \frac{\pi}{l_x} \sqrt{\frac{2}{n\pi}} = \frac{1}{l_x}\sqrt{\frac{2\pi}{n}}
\end{equation}
so that with high probability $a_x$ does not exceed $\pi/l_x$. The $g$-factor difference between ground and excited valley states $\Delta g(x)$ is modelled similarly with
\begin{equation}
    \Delta g(x) = \sum_{k=1}^n \beta_k \sin(\mu_k x)
\end{equation}
where $n=20$ and $\mu_k$ is a random number between $l_x$ and $2dl_{dd}$. A typical maximum value for $\Delta g$ at a constant field of 1T is 100MHz \cite{patomäki2023pipeline}, thus we decide to randomly sample each $\beta_k$ between 0 and $\sqrt{\frac{2}{n\pi}}\times100$MHz.

Finally, in order to understand the phase accumulation over a whole stabiliser cycle, we prepare the valley+spin electron wavefunction in
\begin{equation}
    \ket{\psi_0} = \ket{0} \otimes \ket{+}
\end{equation}
We assume a shuttling speed $v=10$m/s and a dot separation $l_{dd}=140$nm as in the main text. The electron spatial distribution is set to $l_x=50$nm, and we choose $E_{VS,0}=150\mu$eV, so that in the worst-case of $a_xl_x=\pi$, one gets $E_{VS}=15\mu$eV, which in the lowest possible range for $E_{VS}$ \cite{Langrock_2023} (therefore increasing non-adiabatic errors). We aim to estimate the influence of the code size $d$ on the valley-induced dephasing noise, as it controls the overall shuttling distance. Further, we study the effect of additionally flipping the spin state via an $X$ gate at the turning point in the electron trajectory, in the spirit of dynamical decoupling. Indeed, ignoring any disorder, we observe that a flipped spin simply precesses back to its initial state when shuttled back. Hence one may expect that the same phenomenon happens in the presence of disorder and could therefore be used to reduce the noise. The evolution of the wavefunction over time is obtained by solving Schrodinger equation by time discretisation over 50,000 time steps. The final probability of a $Z$ error is given by the overlap of the final state with the $\ket{-}$ spin state. In the presence of random disorder, we estimate the dephasing probability via Monte-Carlo simulations, by repeating the same experiment $N_{reps}=10,000$ times.

\begin{figure}
    \centering
    \includegraphics[width=\linewidth]{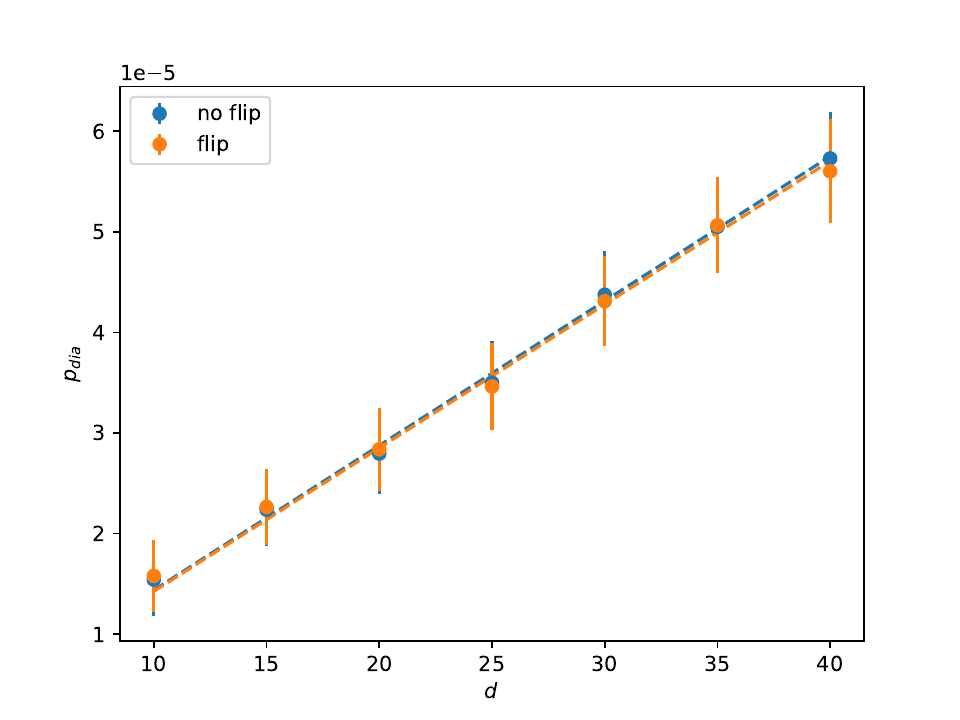}
    \caption{Solid lines: valley-induced dephasing error $p_{dia}$ against the code distance $d$, to which the shuttling distance is proportional. Two cases are plotted, depending on if the spin is flipped via an $X$ before the electron is shuttled back. Dashed lines: linear regression of both cases.}
    \label{fig:valley_induced_dephasing}
\end{figure}

Fig. \ref{fig:valley_induced_dephasing} shows the evolution of the valley-induced dephasing probability $p_{dia}$ against the code distance $d$, both with and without flipping the spin at the turning point of the trajectory. One can see that the error rate is not lowered by such manipulation: we therefore won't execute this operation. By fitting the \textit{no flip} data with a simple linear regression $y=\gamma x$, one can infer a per-dot error probability $p_{dia}/4d=1.4\times 10^{-6}$. As a comparison, the per-dot error induced by adiabatic shuttling for $T_2^*=8\mu$s, \textit{i.e.} in the most optimistic case considered in the main text, is $p_{adia}/4d=4\times 10^{-6}$. This justifies the addition of such non-adiabatic effects in our modelling.

\section{Good matrix for HGP code} \label{appendix:hgp_matrix}

In Sections \ref{sec:hgp} and \ref{sec:Comparative performance}, we simulated the performance of a HGP constructed from the product of a [8,1,8] repetition code and a [17,3,8] classical code generated randomly. Its parity check matrix $H_8$ is
\begin{equation*}
H_8 = \left(
\begin{smallmatrix}
  0 & 1 & 0 & 0 & 0 & 0 & 0 & 0 & 0 & 0 & 0 & 1 & 0 & 1 & 0 & 0 & 0\\
  1 & 0 & 0 & 0 & 1 & 0 & 0 & 1 & 0 & 0 & 0 & 0 & 1 & 0 & 1 & 0 & 0\\
  1 & 0 & 0 & 0 & 0 & 1 & 1 & 0 & 0 & 0 & 0 & 0 & 0 & 0 & 0 & 0 & 0\\
  0 & 0 & 1 & 0 & 0 & 0 & 0 & 0 & 0 & 0 & 0 & 0 & 0 & 0 & 1 & 0 & 0\\
  0 & 0 & 0 & 0 & 0 & 0 & 0 & 0 & 1 & 0 & 0 & 0 & 0 & 1 & 0 & 0 & 0\\
  0 & 0 & 0 & 0 & 0 & 0 & 0 & 1 & 0 & 0 & 0 & 0 & 0 & 0 & 0 & 0 & 0\\
  0 & 0 & 0 & 0 & 1 & 0 & 0 & 0 & 0 & 0 & 0 & 1 & 0 & 0 & 0 & 0 & 0\\
  0 & 0 & 0 & 0 & 0 & 1 & 0 & 0 & 0 & 1 & 0 & 0 & 0 & 0 & 0 & 0 & 0\\
  0 & 0 & 0 & 0 & 0 & 1 & 0 & 0 & 0 & 0 & 0 & 0 & 0 & 0 & 0 & 1 & 0\\
  0 & 0 & 0 & 0 & 1 & 0 & 0 & 0 & 0 & 0 & 0 & 0 & 1 & 0 & 0 & 0 & 1\\
  0 & 0 & 0 & 0 & 0 & 0 & 0 & 0 & 0 & 0 & 1 & 0 & 0 & 0 & 0 & 0 & 0\\
  0 & 1 & 0 & 1 & 1 & 0 & 0 & 0 & 0 & 0 & 0 & 0 & 0 & 0 & 0 & 0 & 0\\
  1 & 1 & 0 & 0 & 0 & 0 & 0 & 0 & 0 & 0 & 0 & 0 & 0 & 0 & 0 & 0 & 0\\
  0 & 0 & 0 & 0 & 0 & 0 & 0 & 0 & 0 & 1 & 0 & 0 & 0 & 1 & 0 & 0 & 1\\
\end{smallmatrix}
\right).
\end{equation*}

In Section \ref{sec:Comparative performance}, we followed the same protocol with a HGP code constructed from the product of a [4,1,4] repetition code and a [7,3,4] classical code generated randomly. Its parity check matrix $H_4$ is
\begin{equation*}
H_4 = \left(
\begin{smallmatrix}
  1 & 0 & 0 & 1 & 1 & 0 & 1\\
  0 & 0 & 1 & 0 & 1 & 0 & 1\\
  0 & 1 & 1 & 0 & 0 & 1 & 0\\
  1 & 0 & 1 & 0 & 1 & 1 & 0\\
\end{smallmatrix}
\right).
\end{equation*}

\end{document}